%% file: manuscript.tex
\documentclass[screen,manuscript,acmsmall]{acmart}
\AtBeginDocument{%
  }

\setcopyright{acmlicensed}
\copyrightyear{2018}
\acmYear{2018}
\acmDOI{XXXXXXX.XXXXXXX}
\acmConference[Conference acronym 'XX]{Make sure to enter the correct
  conference title from your rights confirmation email}{June 03--05,
  2018}{Woodstock, NY}
\acmISBN{978-1-4503-XXXX-X/2018/06}

\usepackage{algorithm}
\usepackage{booktabs}
\usepackage{subcaption}
\usepackage{mwe}
\usepackage[noend]{algpseudocode}

\usepackage{xcolor}
\usepackage{amsmath}

\usepackage{amssymb} 
\usepackage{adjustbox}
\usepackage{array}
\usepackage{mdframed}
\begin{document}

\title{How Millions Coordinate at Scale: Engagement, Collaboration, and Conflict in Three Editions of Reddit r/place}

\author{Yutong Wu}

\email{yw180@rice.com}
\affiliation{%
  \institution{Rice University}
  \city{Houston}
  \state{Texas}
  \country{USA}
}

\author{Arlei Silva}
\email{arlei@rice.com}
\affiliation{%
  \institution{Rice University}
  \city{Houston}
  \state{Texas}
  \country{USA}
}


\begin{abstract}

Mass peer-production environments are shaped by a complex interplay between decentralized coordination, platform design, and potential conflict over resources. While online infrastructures enable large-scale collaboration, they can also introduce coordination challenges and contested interactions among participants. The Reddit r/place experiment provides a unique socio-technical setting for studying this type of dynamics across three distinct editions (2017, 2022, and 2023). By allowing millions of participants to collaborate and compete as they update pixels on a finite shared digital canvas, the events generated fine-grained traces comprising hundreds of millions of actions taken over multiple days.

In this paper, we examine how participation, collaboration, and conflict evolve within r/place. First, we conduct a longitudinal, cross-edition analysis of engagement and collaboration patterns across the 2017, 2022, and 2023 events. Second, to investigate collaborative activity beyond surviving final artifacts, we introduce a scalable graph-based dynamic clustering framework and apply it to the 2017 event to reconstruct coalition trajectories from behavioral interaction logs, enabling the analysis of both persistent and transient efforts.

Our findings reveal recurring organizational patterns across editions: participation remains highly concentrated to few participants despite individual rate-limiting constraints, larger coalitions exhibit both greater coordination inefficiencies and lower median per-participant activity, while still success increasingly concentrates within large collaborative groups over the course of an event. Analysis of recovered coalition trajectories further shows that coalition outcomes are difficult to predict based on state characteristics during much of the event, highlighting the dynamic and contested nature of collaborative production in r/place.

These findings provide new insight into the lifecycle of large-scale online collaboration and suggest design considerations for collaborative systems that must balance participation, coordination, and competition.

\textbf{Content Warning:} \textit{This paper includes direct quotes that contain graphic language}.
\end{abstract}



\maketitle

\input{introduction}

\input{related_work}
\input{datasets}
\input{engagement}
\input{collaboration}
\input{competition}

\input{discussion}
\bibliographystyle{ACM-Reference-Format}
\bibliography{sample-base}

\appendix
\input{appendix}

\end{document}

%% file: introduction.tex
\section{Introduction}

``\textit{There is an empty canvas. You may place a tile upon it, but you must wait to place another. Individually you can create something. Together you can create something more}''. This is how the first Reddit r/place experiment was announced in 2017 \cite{simpson-how}. Since then, r/place has evolved into a massive peer production phenomenon, drawing over 100 million participants across three distinct editions \cite{cuthbertson-place, lorenz-place}. The experiment operates under strict socio-technical constraints: individuals can alter a single pixel's color on a shared digital canvas but are restricted by a five-minute cooldown timer between actions \cite{simpson-how}. To overcome these constraints and build lasting digital artifacts, users must engage in massive, distributed crowd coordination. However, this environment is inherently adversarial; space on the canvas is finite, and less than 7\% of individual pixel updates survive to the final snapshot. This unique dynamic creates an unprecedented socio-technical laboratory to study how platform infrastructure, temporal bursts, and emergent user-built tooling shape the lifecycles of large-scale cooperative-competitive crowd initiatives \cite{benkler2002coase,benkler2006wealth,rafaeli2008online,forte2008scaling,forte2009decentralization}.


While the previous literature explores peer production within frameworks, such as Wikipedia \cite{rafaeli2008online, forte2008scaling}, citizen science initiatives \cite{mason2012collaborative}, or gaming guilds \cite{chen2008player, jagannath2020we}, r/place introduces distinct architectural constraints and adversarial pressures that challenge traditional collaborative paradigms. Any pixel placement can instantaneously overwrite a predecessor's output. This zero-sum environment introduces dual layers of systemic friction: explicit platform constraints (such as expanding canvas boundaries and temporal rate limits) and highly volatile socio-technical group dynamics (including participation patterns, escalating coordination inefficiencies, and outcome unpredictability). Over the course of three editions (2017, 2022, and 2023), participant collectives adapted to these challenging conditions, evolving from disorganized individual actions into ad-hoc community alliances, and ultimately into highly structured, multi-platform defensive coalitions. These networks dynamically integrated external communication hubs and automated placement scripts to safeguard their spatial real estate on the canvas. Our empirical understanding of how such massive, decentralized crowd coalitions emerge, scale, and maintain stability against inter-group conflicts and uncertainty remains limited. Gaining insight into these mechanics is vital for understanding how platform architecture and group dynamics co-evolve to actively shape collective action in mass resource-constrained online peer production.

To evaluate these multi-edition dynamics, we trace individual canvas updates as \textit{actions}, contributing users as \textit{participants}, and groups collaborating on a shared canvas artifact as a \textit{coalition}. Under this framework, our work examines collective action through three complementary dimensions: engagement, collaboration, and conflict. Engagement concerns the relationship between platform characteristics, user interests, and voluntary participation over time. Collaboration addresses how decentralized coalitions organize and distribute labor under explicit rate-limiting constraints. Conflict examines how adversarial spatial interactions and outcome unpredictability are associated with the persistence of peer-produced artifacts. Whereas previous work on r/place relies on human-curated atlases that document successful final artifacts \cite{vachher2020understanding,israeli2022must,israeli2023flying}, our study seeks to incorporate the histories of transient collaborations through dynamic coalition recovery.

Our paper aims to enhance our understanding of mass crowd production under inter-group conflict by addressing three key research questions using r/place: 
\begin{enumerate}
\item \textbf{RQ1 (Engagement):} How does participant engagement vary across platform scales, temporal bursts, user interests, and structural constraints over distinct editions?

\item \textbf{RQ2 (Collaboration):} How do participation patterns and coordination inefficiencies vary across coalitions of different sizes?

\item \textbf{RQ3 (Conflict):} How do conflict dynamics relate to coalition persistence, artifact stability, and outcome predictability?
\end{enumerate}

We address these questions by conducting a cross-edition quantitative analysis across three iterations of r/place, pairing detailed behavioral logs of hundreds of millions of pixel actions with contextual community data collected via the Reddit API \cite{simpson-how}. To address the survival bias inherent in static canvas snapshots, we develop a scalable, dynamic clustering framework. By combining graph-based image segmentation with representation learning, our method successfully reconstructs the developmental lifecycles of both visible and historically erased labor from coalitions. 

The contributions of this paper are threefold: (1) a longitudinal, cross-edition characterization of engagement, collaboration, and conflict dynamics in the r/place ecosystem; (2) a scalable graph-based clustering framework for recovering and tracking transient and failed coalitions from large-scale behavioral logs; and (3) an empirical analysis of participation patterns, coordination inefficiencies, and adversarial interactions, revealing how these dynamics vary across coalition scales and competitive environments.

\begin{figure*}[t!]
    
    \centering
    \begin{subfigure}[t]{0.27\textwidth}
        \centering
        \includegraphics[width=\linewidth]{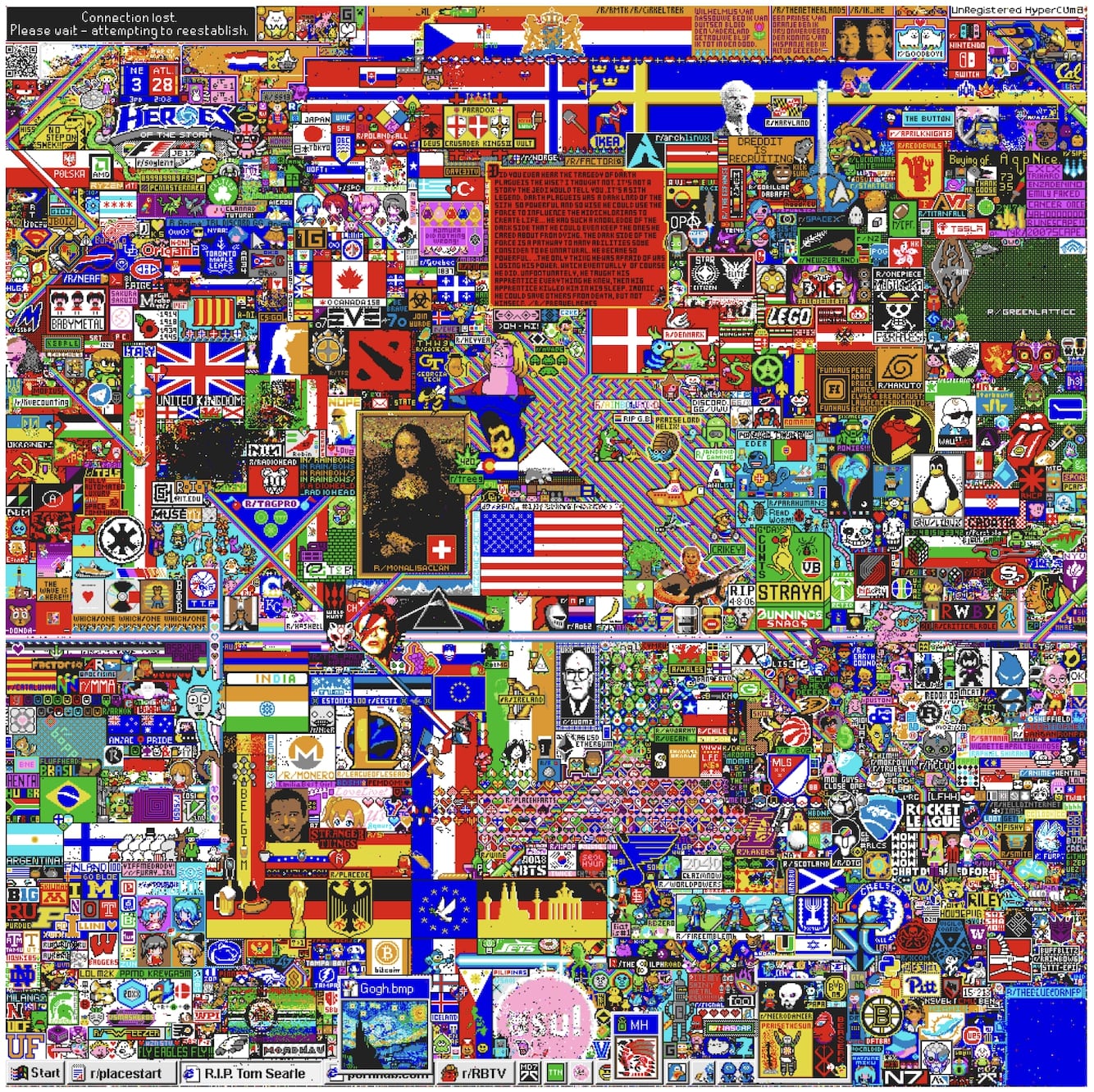}
        \caption{2017}
    \end{subfigure}~
    \begin{subfigure}[t]{0.275\textwidth}
        \centering
        \includegraphics[width=\linewidth]{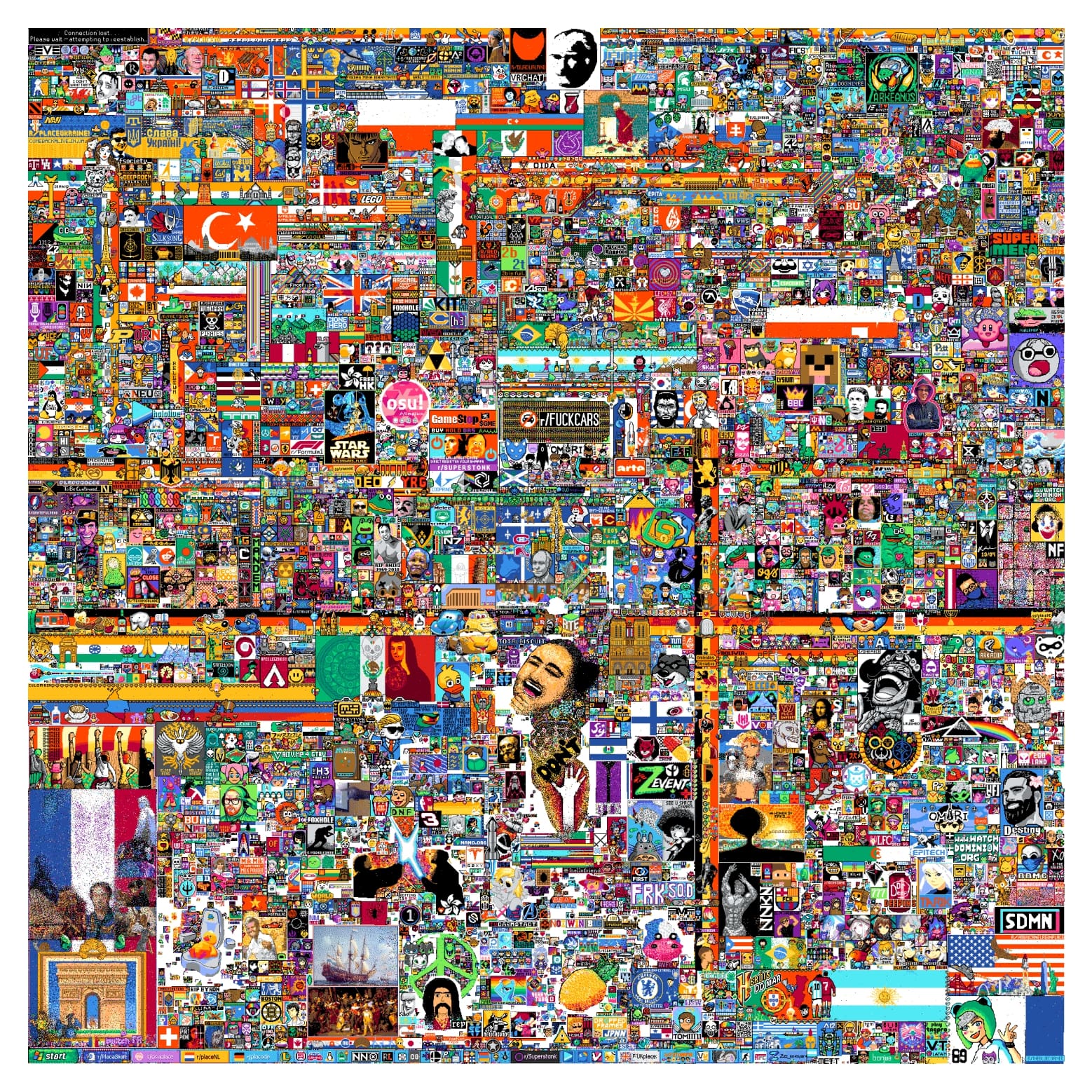}
        \caption{2022}
    \end{subfigure}~
    \begin{subfigure}[t]{0.4\textwidth}
        \centering
        \includegraphics[width=\linewidth]{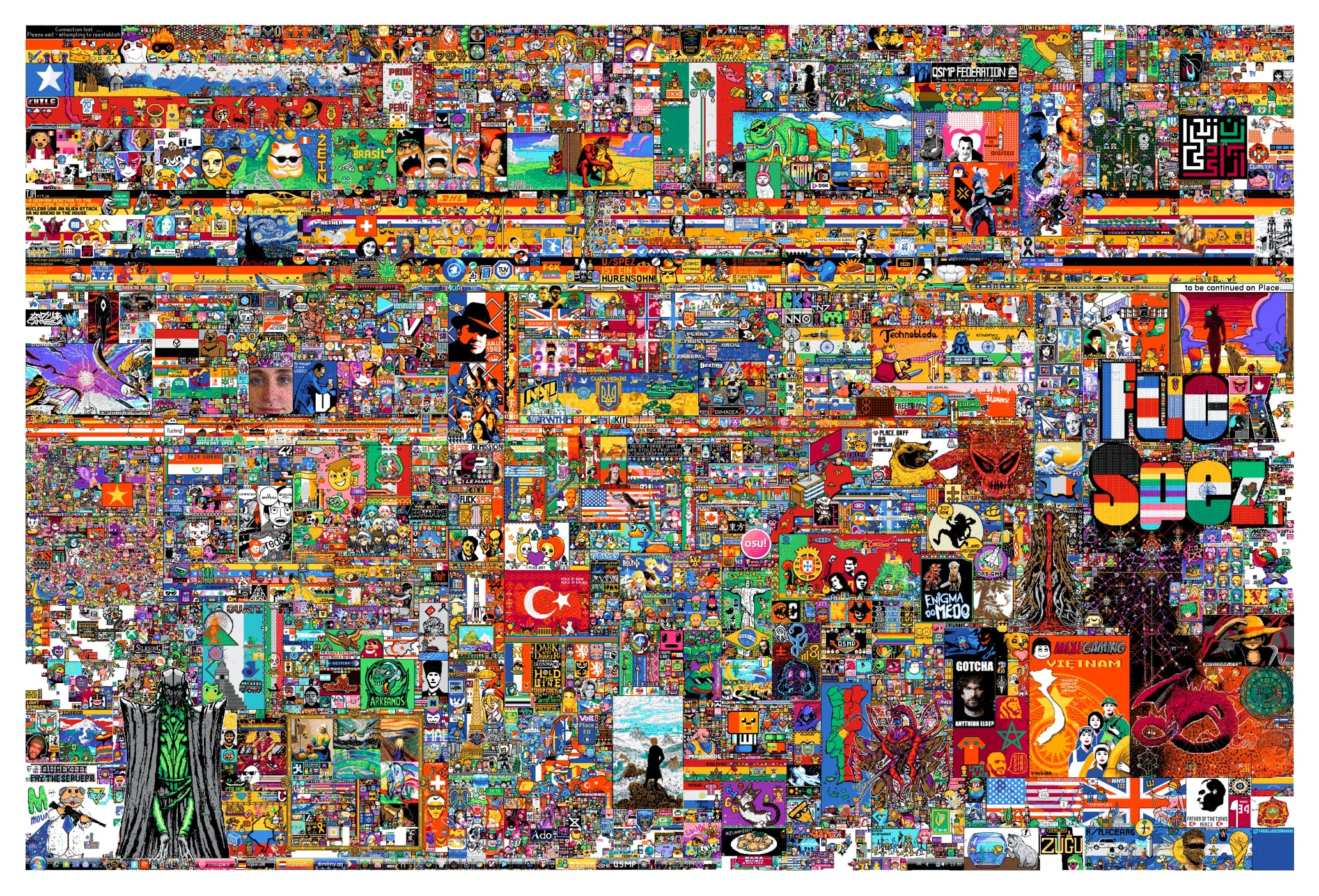}
        \caption{2023}
    \end{subfigure}
    
    \caption{Final canvas for the 2017, 2022, and 2023 editions of r/place (excluding whiteout updates for 2022 and 2023). We leverage r/place to study how platform infrastructure, temporal bursts, and emergent user-built tooling shape the lifecycles of large-scale cooperative and competitive crowd initiatives}
    \label{fig:heatmap}
\end{figure*}

%% file: related_work.tex
\section{Related Work}
\label{sec:related_work}

Our work sits at the intersection of mass peer production, socio-technical systems, and computational social science. We situate our analysis of Reddit r/place within a broader body of research on large-scale online collaboration, collective action, and the organizational dynamics of digital communities \cite{postmes2002collective,raban2010empirical,turner2005picturing,iriberri2009life}. In particular, we draw on prior work examining coordination in peer-production systems, the role of socio-technical infrastructures in shaping collaborative behavior, and computational approaches for identifying and tracking emergent groups \cite{vachher2020understanding,kittur2008harnessing,marlow2013impression}. By studying both successful and transient collaborations across three editions of r/place, our work extends existing research beyond final-state artifacts to examine the full lifecycle of collective production in a competitive, resource-constrained environment.

\subsection{Mass Peer Production and Distributed Workspace Scaffolding}

Mass peer-production systems rely on large, decentralized collectives that coordinate contributions toward shared artifacts \cite{simpson-how, borra2015societal}. Prior research in CSCW has extensively examined collaborative production in environments such as Wikipedia, open-source software projects, and citizen-science platforms, highlighting how socio-technical infrastructures enable coordination, governance, and large-scale knowledge creation \cite{forte2009decentralization, rafaeli2008online, mason2012collaborative, gupta2022instagram}. These systems typically support persistent records, rich communication channels, and mechanisms for revising or branching contributions over time.

Collaborative canvas environments such as r/place introduce a substantially different coordination context. Rather than supporting parallel development or version-controlled revision, the shared canvas constitutes a finite and highly contested workspace in which every action directly modifies a common artifact. Because pixel updates overwrite previous contributions, collaboration and competition become tightly intertwined, requiring participants to simultaneously coordinate within groups and respond to actions from competing collectives.

These characteristics connect r/place to broader CSCW discussions surrounding coordination challenges in large-scale collaboration \cite{cummings_07, Pendharkar_09,anand2023game}. Prior work has shown that coordinating distributed labor becomes increasingly difficult as groups grow, often resulting in higher communication overhead, reduced situational awareness, and uneven participation \cite{cheng2019makes, mason2012collaborative}. At the same time, research on collective intelligence and peer production has emphasized the importance of organizational structures and socio-technical scaffolds in supporting large-scale collaborative efforts \cite{malone2010collective, rand2013human}.

Building on this literature, our study examines how participation patterns, coordination inefficiencies, and conflict dynamics vary across coalition scales and platform editions. Rather than focusing exclusively on successful final artifacts, we investigate the full lifecycle of collaborative efforts—including transient and unsuccessful coalitions—allowing us to characterize how large-scale coordination unfolds within a competitive and resource-constrained workspace.

\subsection{Socio-Technical Conflict, Resistance, and Platform Governance}

Conflict is as a recurring feature of large-scale online collaboration, particularly in environments where participants compete over shared resources, visibility, or control of collective artifacts \cite{vachher2020understanding,kriplean2007community}. Research in social computing and social psychology has shown that adversarial interactions can reinforce group boundaries, strengthen collective identities, and influence patterns of participation and coordination \cite{tajfel2001integrative}. More broadly, theories of inter-group competition suggest that competitive environments may encourage greater levels of within-group organization and cooperation \cite{richerson2016cultural}.

Prior studies have examined these dynamics across a range of collaborative platforms. In text-based environments such as Wikipedia, conflict often emerges through edit wars, content disputes, and governance disagreements, which are typically managed through moderation policies, access controls, and community oversight mechanisms \cite{borra2015societal,kriplean2007community}. These governance structures provide formal mechanisms for resolving disputes and protecting contested contributions.

Canvas environments present a different coordination context. Because contributions directly overwrite one another on a shared and finite workspace, maintaining an artifact requires sustained collective effort rather than administrative intervention alone. Prior work on r/place has documented how participants organized through external communication channels and community infrastructures to coordinate activity and defend shared artifacts \cite{vachher2020understanding,israeli2022must,israeli2023flying}. These studies highlight the importance of socio-technical scaffolding—including communication platforms, overlays, and automation tools—in supporting collective action within highly contested digital spaces.

Building on this literature, our work examines how conflict dynamics relate to coalition formation, persistence, and outcomes across multiple editions of r/place. By recovering both successful and transient collaborations, we extend prior analyses beyond surviving artifacts to investigate how competitive pressures shape the broader lifecycle of peer-production efforts.

\subsection{Empirical Studies of Reddit r/place}

Since its introduction, r/place has become a valuable empirical setting for studying large-scale collective behavior, online coordination, and emergent social organization \cite{rappaz-place, chen2021collaborative, israeli2022must, pendergrass2022digital, litherland2021instruction, muller2018compression, armstrong2018coordination, vachher2020understanding}. Prior work has examined a diverse range of phenomena, including the spatial organization of cultural communities \cite{muller2018compression}, the relationship between canvas activity and external Reddit discussions \cite{israeli2022must}, and the dynamics of territorial conflict among participating groups \cite{vachher2020understanding}. Collectively, these studies demonstrate how r/place serves as a rich socio-technical environment for investigating large-scale collaborative and competitive behavior.

Despite these advances, much of the existing literature focuses on either final-state artifacts or a limited subset of highly visible communities. As a result, less is known about the lifecycle of transient collaborations, unsuccessful initiatives, and the broader organizational dynamics that emerge throughout the event. Our work builds on this literature through a longitudinal analysis spanning three editions of r/place and a coalition recovery framework designed to identify both persistent and transient collaborative efforts. This perspective enables us to examine how participation, coordination, and conflict evolve over time, including collaborative activity that is not captured by final canvas configurations.

\subsection{Temporal Segmentation and Recovering Transient Collaboration}

Understanding conflict and coordination in large-scale collaborative systems requires transforming low-level behavioral traces into meaningful socio-technical units. This challenge is closely related to community detection in complex networks \cite{fortunato2010community,newman2006modularity} and spatio-temporal clustering in dynamic data \cite{jain1999data,ward1963hierarchical}. Within the r/place literature, the approach most closely aligned with this objective is that of Rappaz et al. \cite{rappaz-place}, who proposed a representation-learning framework for grouping participant actions using a Bayesian Personalized Ranking (BPR) objective. However, existing approaches are primarily evaluated against final canvas states or community-curated atlases, which predominantly capture artifacts that remain visible at the end of the event.

Our work extends this line of research by introducing a scalable dynamic clustering framework that integrates image segmentation \cite{Felzenswalb-graphcut} with graph-based participant representations \cite{grover2016node2vec}. By linking artifact partitions across temporal snapshots through a set-cover-inspired matching procedure, the framework reconstructs evolving coalition trajectories over time. This approach enables the analysis of both persistent and transient collaborative efforts, providing a broader view of participation, coordination, and conflict than is possible from final-state artifacts alone. Consequently, it offers a richer empirical foundation for studying coalition lifecycles and organizational dynamics in large-scale peer-production environments.

%% file: datasets.tex
\section{Behavioral and Community Datasets}
This section summarizes the three Reddit r/place experiments and describes the multi-edition behavioral and community datasets used in our analysis. Figure \ref{fig:heatmap} shows the final snapshot canvas for each edition of the r/place experiment.

\subsection{Reddit r/place}
\label{sec::r-place}
The r/place experiments were massive-scale peer production events hosted by Reddit in 2017, 2022, and 2023. While the micro-level mechanics remained constant (i.e., individual participants altering single pixel colors on a shared public canvas, subjected to an architectural five-minute rate limit \cite{simpson-how}), each event introduced distinct structural affordances, governance mechanisms, and macro-social contexts. Table \ref{table:data_highlight} summarizes the experiment statistics across the three editions.

\textbf{2017 edition:} Operating for 72 hours on a fixed, static $1000 \times 1000$ pixel canvas with a 16-color palette, the inaugural edition recorded 16,559,897 actions from 1,166,925 unique accounts. This baseline event marked the transition from ad-hoc, small-group drawings to structured, multi-platform coordination as users coordinated their efforts via subreddits \cite{cuthbertson-place} and deployed open-source automation scripts (e.g., placement bots) to expand and protect digital artifacts against territorial competition and vandalism \cite{majd-placestart, vachher2020understanding}.

\textbf{2022 edition:} Spanning 87 hours, the second iteration introduced an expanding spatial environment that dynamically doubled the canvas footprint over time ($1000 \times 1000 \to 1000 \times 2000$ at hour 27, and $2000 \times 2000$ at hour 54), paired with a multi-tiered, changing color palette (16, 24, and 32 colors). To manage an explosive traffic surge to 108,034,224 accounts and 149,560,838 updates, the event featured updated moderation filters and bot-containment protocols. This edition was defined by high-intensity conflict as decentralized, community-driven micro-coalitions actively defended their digital artifacts against coordinated territorial raids at times orchestrated by high-profile livestreamers and bot networks \cite{lorenz-place, eudaly-fans}. The event closed with a platform-enforced ``whiteout'' phase, restricting input solely to white pixels to clean the canvas.

\textbf{2023 edition:} Running for 125 hours, the final edition featured an accelerated, six-tiered canvas expansion scaling up to a $3000 \times 2000$ layout. Recording 122,719,796 actions from 78,582,675 participants, this edition experienced a contraction in individual participation alongside an escalation in automated labor and platform dissent. Occurring shortly after sweeping modifications to Reddit's API infrastructure pricing, the socio-technical ecosystem became deeply politicized \cite{jay-place, jay2-place, jay3-place}. Massive user coalitions weaponized the canvas mechanics to execute distributed protests against platform governance, embedding prominent anti-corporate messaging (e.g., ``Fuck Spez'') directly into the canvas despite Reddit's efforts to contain hate speech in the experiment \cite{jody-removal}.

\subsection{Behavioral Dataset}
Our primary behavioral dataset maps the complete historical activity trace of the canvas, capturing every pixel update defined as a tuple of the account identifier, spatial coordinates, color, and timestamp. To extract organizational meaning from these unstructured pixel logs, we build a mapping framework using community-driven bounding box annotations compiled via the open-source \textit{r/place Atlas} project \cite{stefano-atlas}. The atlas provides human-annotated polygon boundaries mapping final canvas locations to specific subreddit names, project descriptions, and project URLs. We manually audited the atlas to ensure data integrity. To trace individual pixels to collaborative groups, we model the assignment as a spatial point-in-polygon assignment problem in computational geometry, resolved using a ray-casting algorithm \cite{Shimrat1962}. 

\subsection{Community Metadata}
Because coordination on the canvas was structured externally, we complement our behavioral data with communication traces extracted directly from the text-based Reddit platform. Using the Reddit API, we collected all posts and comments across participating subreddits before, during, and
after each edition of r/place. More specifically, our analysis will be based on subreddit data from
March 1 to April 30 in 2017 and 2022, and from July 1 to August 31 in 2023.

\begin{table}[ht!]
\begin{adjustbox}{width=1\textwidth}
    \begin{tabular}{ |c|c|c|c| } 
    \hline
     \textbf{Year} & \textbf{2017} & \textbf{2022} & \textbf{2023} \\
     \hline
     Colors & 16 & 16$\to$24$\to$32$\to$1 & 8$\to$16$\to$24$\to$32$\to$1\\
     \hline
     Duration (in hours) & 89 & 81 & 125\\
     \hline
     Canvas size & 1000 x 1000 & 1000 x 1000$\to$2000 x 2000 & 1000 x 1000$\to$3000 x 2000\\
     \hline
     Number of actions & 16,559,897 & 149,560,838 & 122,719,796\\ 
     \hline
     Number of participants & 1,166,925  & 108,034,224 &78,582,675\\ 
     \hline
     Labeled artifacts on the final snapshot & 1,588  &  10,885 & 6,244\\ 
     \hline
     Average area of an artifact & 629.72 & 367.48 & 960.92\\ 
     \hline
     Maximum area of an artifact & 88,281
     & 124,500
     & 1,237,932 
     \\
     \hline
     Average actions per player & 14.19 & 1.38 & 1.56\\
     \hline
     Maximum actions per player & 545  &693 & 3904\\
     \hline
     Average coalition size & 734.84  & 9,925.06 & 12,585.31\\
     \hline
     Maximum coalition size & 54,510  & 1,068,986& 4,777,294\\
     \hline
    Potential bots & 235 (0.02\%) & 1,457 (0.00001\%)& 161,640 (0.21\%)\\
     \hline
          Actions by potential bots & 51,953 (0.3\%) & 274,579 (0.002\%)& 20,938,176  (17.1\%)\\
     \hline
    \end{tabular}
    \end{adjustbox}

    \caption{Statistics of the three datasets (2017, 2022, and 2023).}
    \label{table:data_highlight}
\end{table}

%% file: engagement.tex
\section{Engagement Dynamics (RQ1)}
\label{sec:engagement}

The engagement dynamics of Reddit r/place emerge from a combination of decentralized participation, asynchronous coordination, and platform-imposed constraints. Unlike many collaborative projects that rely on formal task assignment or scheduled participation, r/place allows users to contribute voluntarily at any time while limiting activity through a fixed cooldown interval between actions. This design creates a large-scale environment in which participation is self-organized yet resource-constrained. To characterize these engagement patterns, we analyze behavioral traces from canvas action logs alongside temporal event progression and community-level metadata, examining how participation evolves across editions, communities, and stages of the event.

\subsection{Temporal Distributions and Bot Filtering}

\begin{figure}[b!]
    \centering
    \begin{subfigure}[t]{0.48\textwidth}
        \centering
        \includegraphics[width=\textwidth]{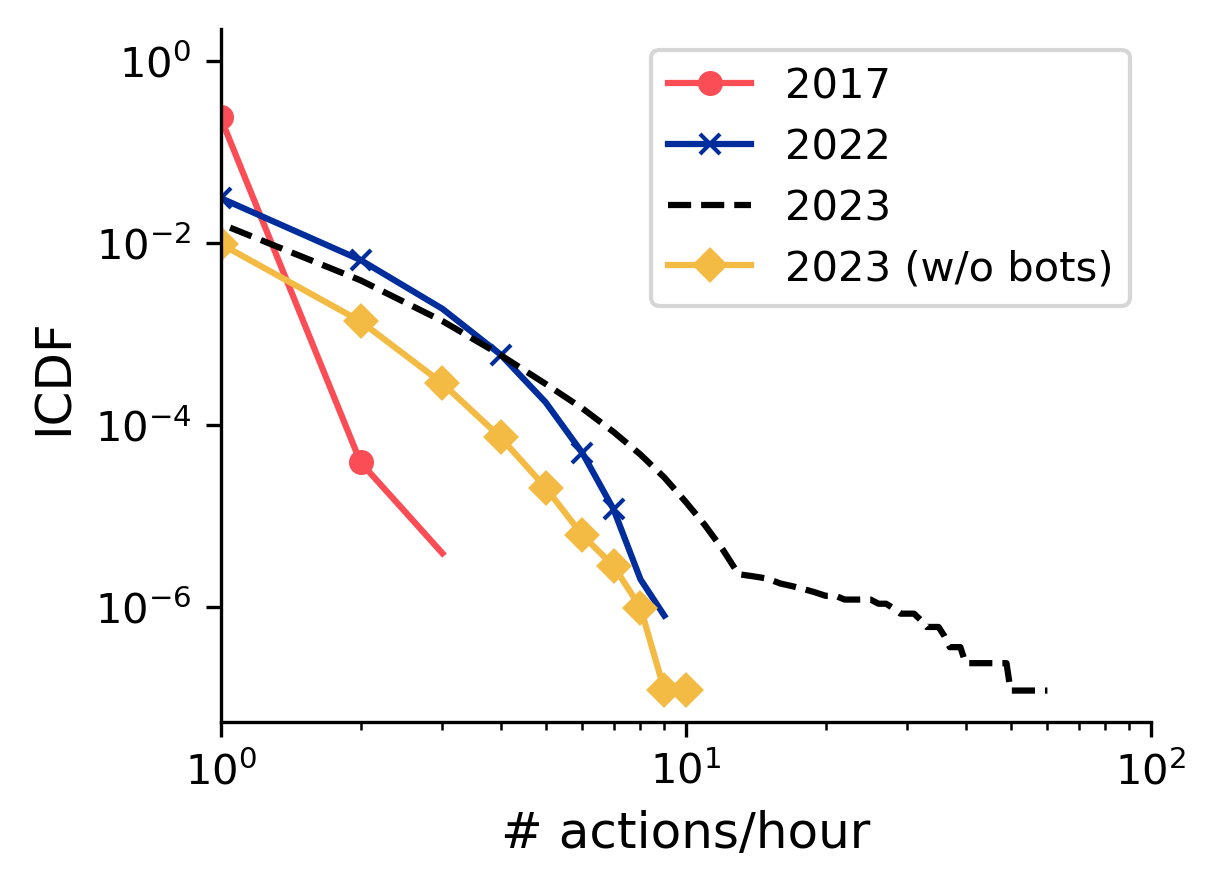}
        \caption{Action distribution rate per participant}\label{fig::updates_hour_user}
    \end{subfigure}\hfill
    \begin{subfigure}[t]{0.48\textwidth}
          \centering
          \includegraphics[width=\linewidth]{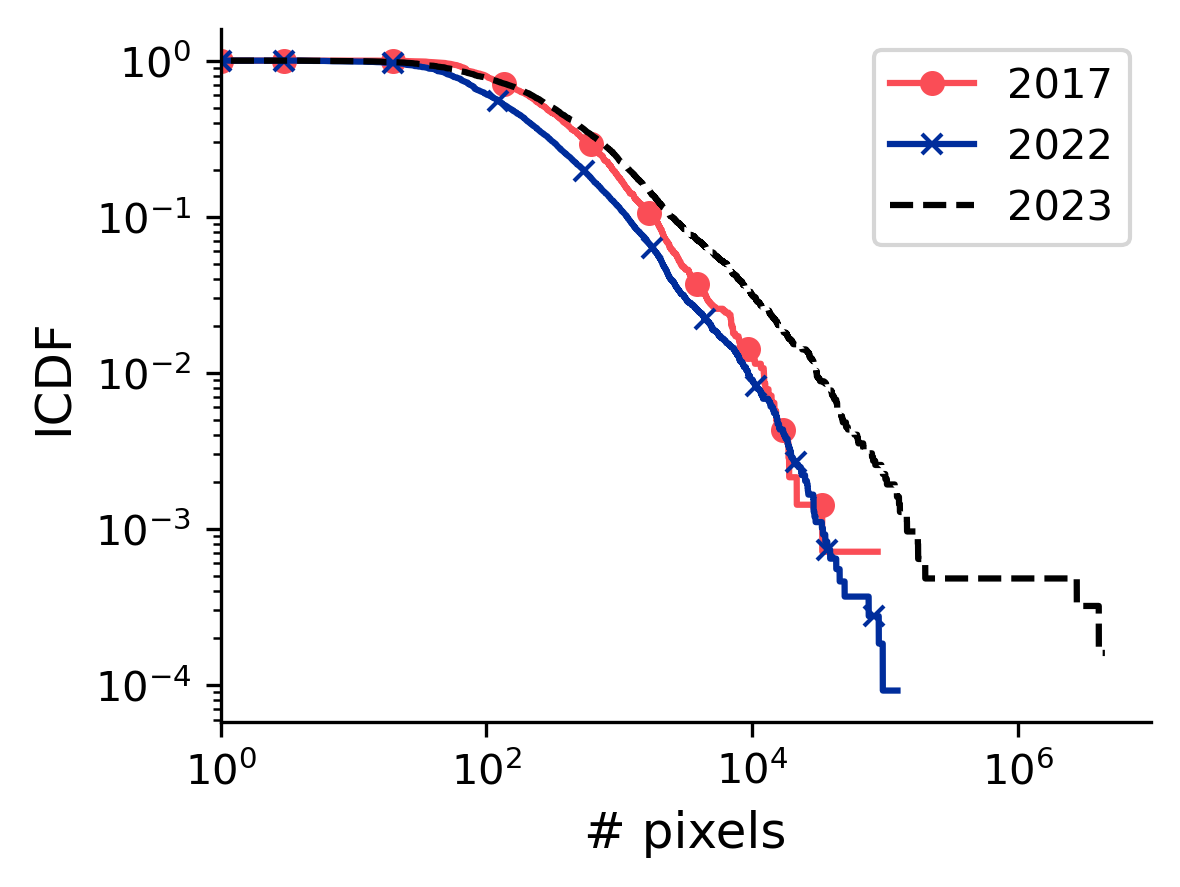}
          \caption{Final retained actions per coalition}\label{fig::updates_drawing}
    \end{subfigure}
    \caption{Inverse cumulative distributions (ICDF) highlighting cross-edition participation scale. Final actions reflect pixels that successfully persisted to the closing canvas snapshot without being overwritten.}
\end{figure}

To characterize participation within this rate-limited environment, Figure \ref{fig::updates_hour_user} presents the Inverse Cumulative Distribution Function (ICDF) of hourly action rates across all three editions. Activity exhibits a pronounced heavy-tailed distribution, with a small fraction of accounts contributing substantially more actions than the typical participant. This pattern is particularly evident in 2023, where the most active identifier recorded 3,904 actions (an average of 31.23 actions per hour).

A subset of accounts in the 2023 dataset exhibits activity levels that substantially exceed the nominal five-minute cooldown interval. While such behavior may arise from automation, coordinated scripting, or other non-standard participation mechanisms, it is unlikely to reflect typical user activity. To focus on participation patterns that are more representative of ordinary users, we exclude accounts exceeding an average rate of one action every five minutes in 2023. Applying the same filter to the 2017 and 2022 datasets identifies only a negligible fraction of accounts (Table \ref{table:data_highlight}), suggesting that unusually high-frequency activity was substantially more prevalent in 2023.

\begin{figure*}[t!]
    \centering
    \begin{subfigure}[t]{0.33\textwidth}
        \centering
        \includegraphics[width=\linewidth]{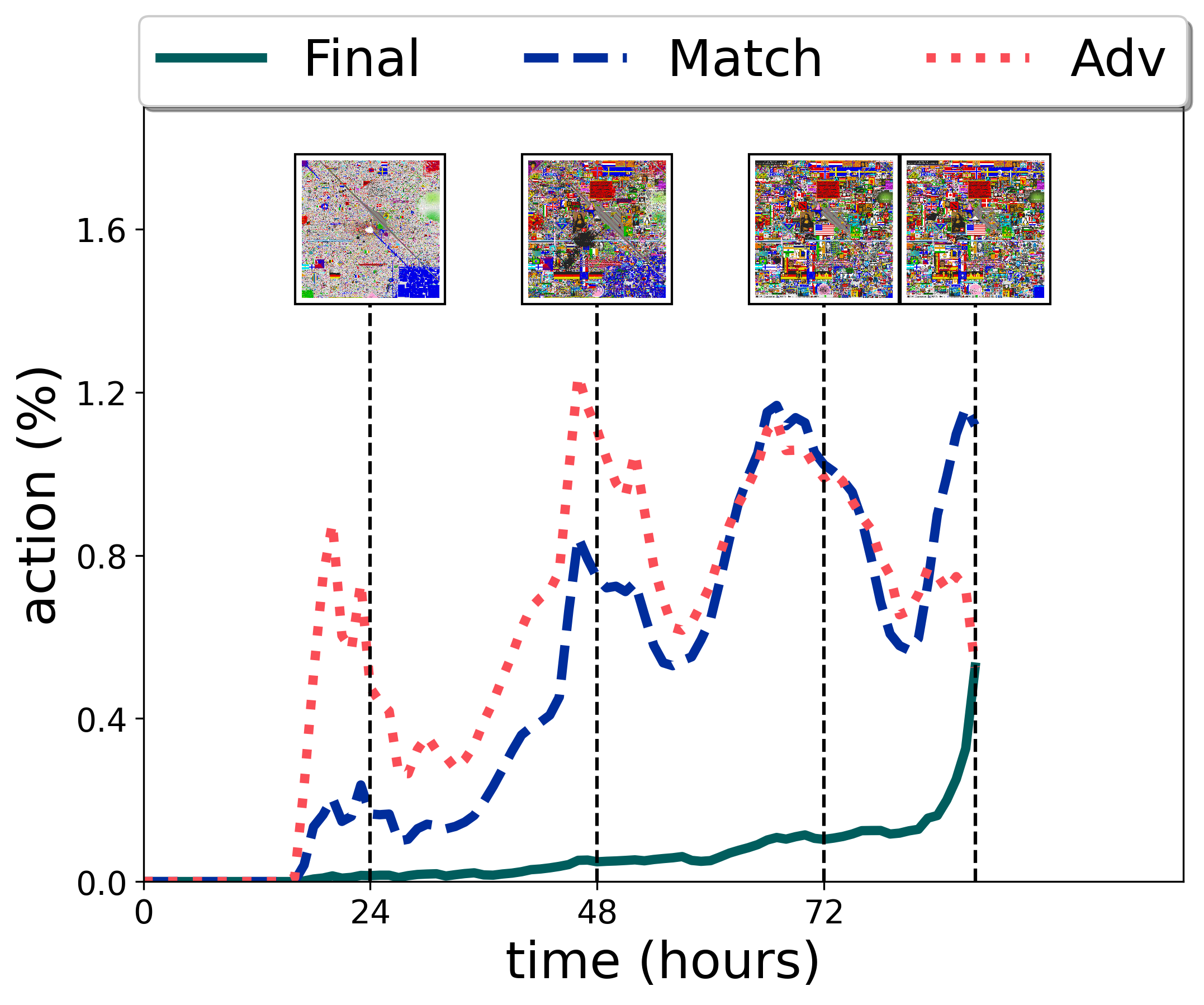}
        \caption{2017 Progression}
    \end{subfigure}~
    \begin{subfigure}[t]{0.33\textwidth}
        \centering
        \includegraphics[width=\linewidth]{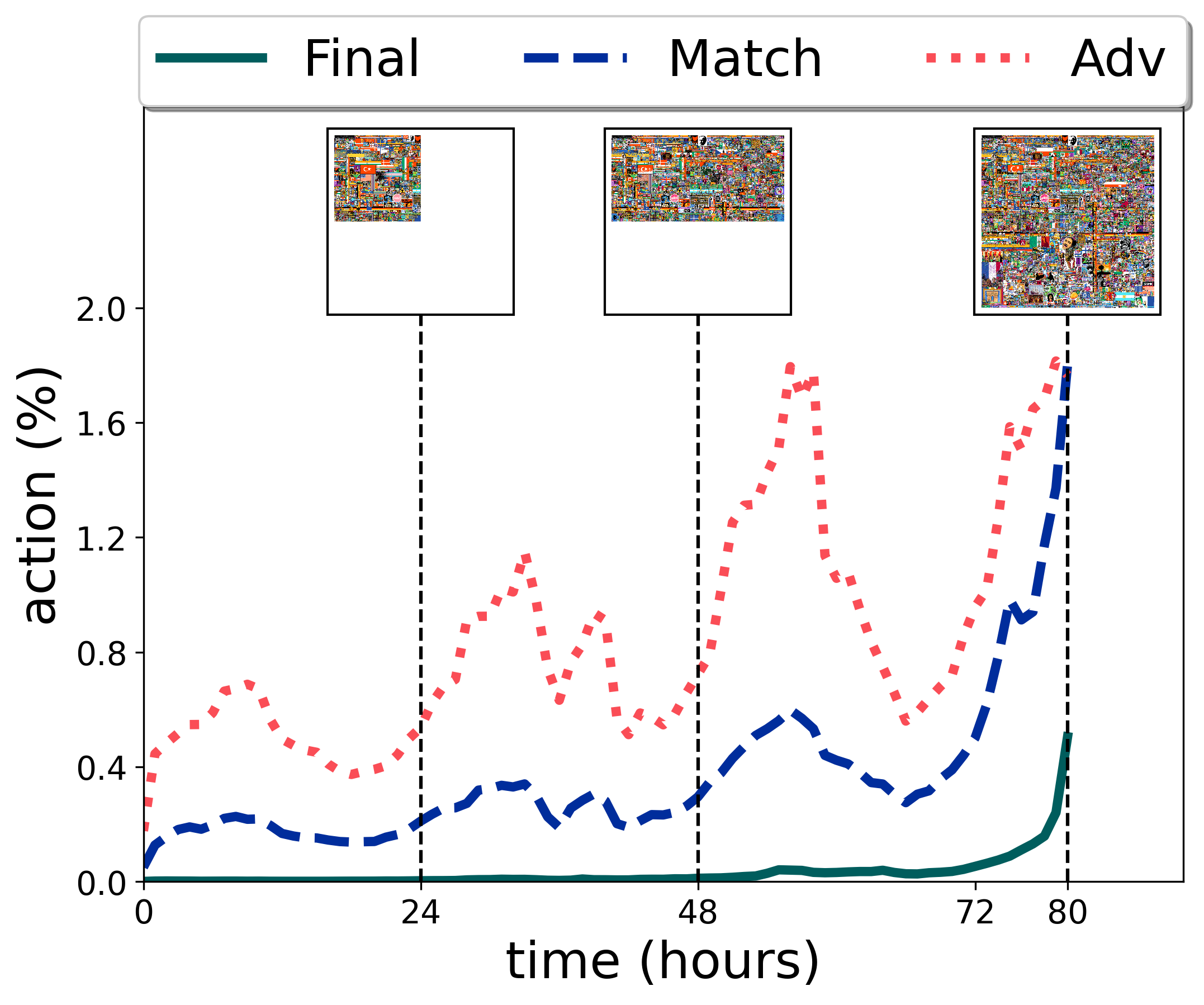}
        \caption{2022 Progression}
    \end{subfigure}~
    \begin{subfigure}[t]{0.33\textwidth}
        \centering
        \includegraphics[width=\linewidth]{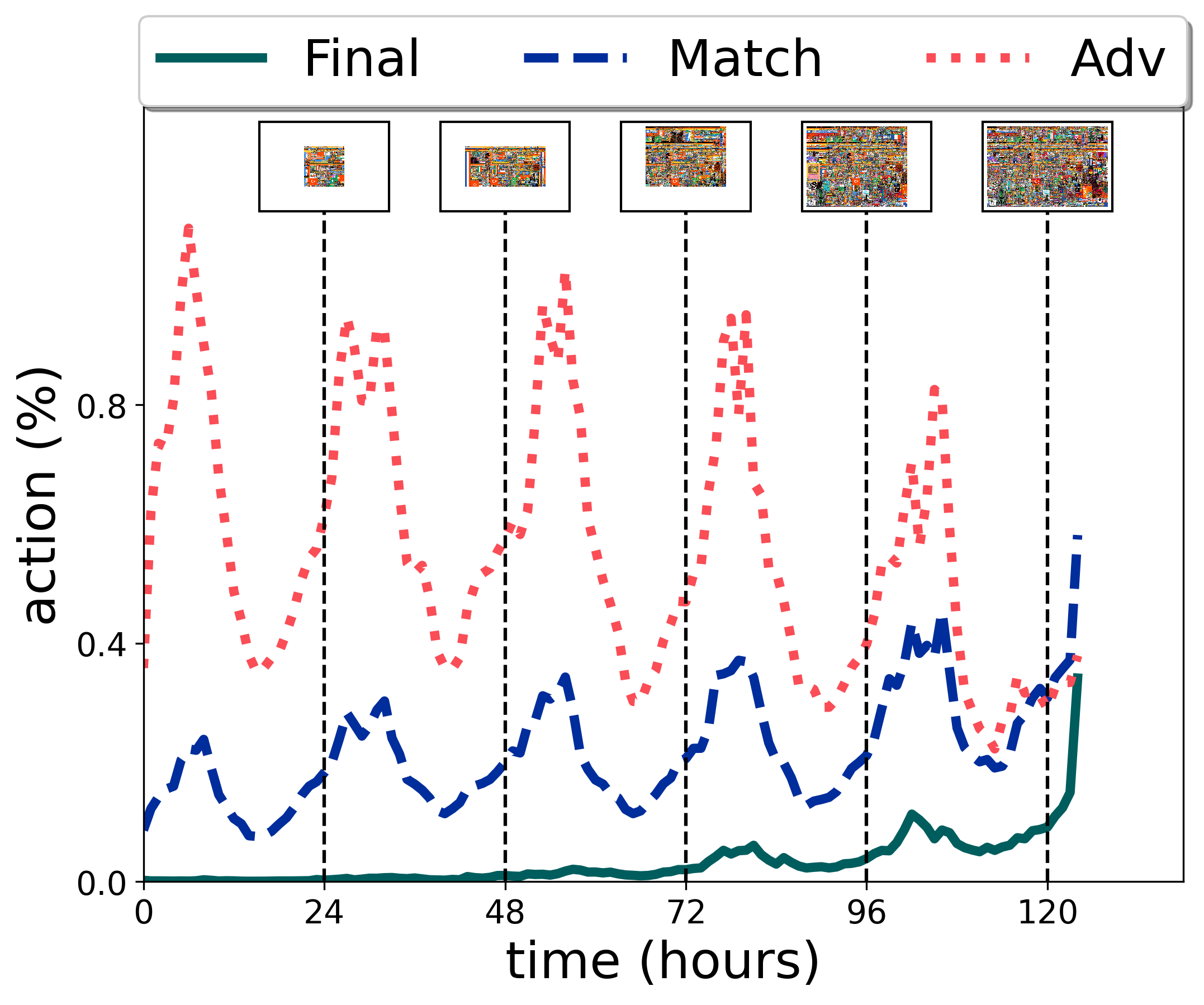}
        \caption{2023 Progression}
    \end{subfigure}
    \caption{Socio-technical canvas progression tracking the composition of active labor. Data represents the temporal percentage of overall actions, distinguishing between permanent (\textit{Final}), transient (\textit{Match}), and conflicting (\textit{Adv}) interactions.}
    \label{fig:canvas_progression}
\end{figure*}

Beyond individual activity rates, participation also varies systematically throughout the lifespan of each event. Figure \ref{fig:canvas_progression} categorizes actions according to their relationship with the final canvas state: \textit{Final} (updates that persist until the closing snapshot), \textit{Match} (updates whose color temporarily aligns with the final artifact but were later overwritten), and \textit{Adversarial} (updates whose color differs from the final artifact). Across all editions, activity follows pronounced diurnal cycles, with participation peaks broadly aligning with evening hours in the Americas. The temporal composition of actions also evolves over the course of the event. Actions associated with the final canvas state become increasingly concentrated toward later stages, whereas much of the event is characterized by ongoing overwrites and competing modifications. While the 2017 and 2022 editions exhibit sustained growth in activity over time, the 2023 edition displays a comparatively flatter temporal profile. This difference is consistent with broader changes in participation dynamics during the 2023 event, although multiple factors may contribute to the observed pattern.

\subsection{Thematic Orientations and Collective Identity Alignments}

To better understand which types of communities were most actively represented on the canvas, we map artifact locations in the 2017 atlas to their associated subreddits. Subreddits were categorized using an LLM-assisted thematic taxonomy (GPT-4o) followed by human auditing and refinement. Communities created primarily for participation in r/place were classified as \textit{``r/place only''}.

\begin{figure}[t!]
    \centering
    \begin{subfigure}[t]{0.48\textwidth}
        \centering
        \includegraphics[width=0.95\textwidth]{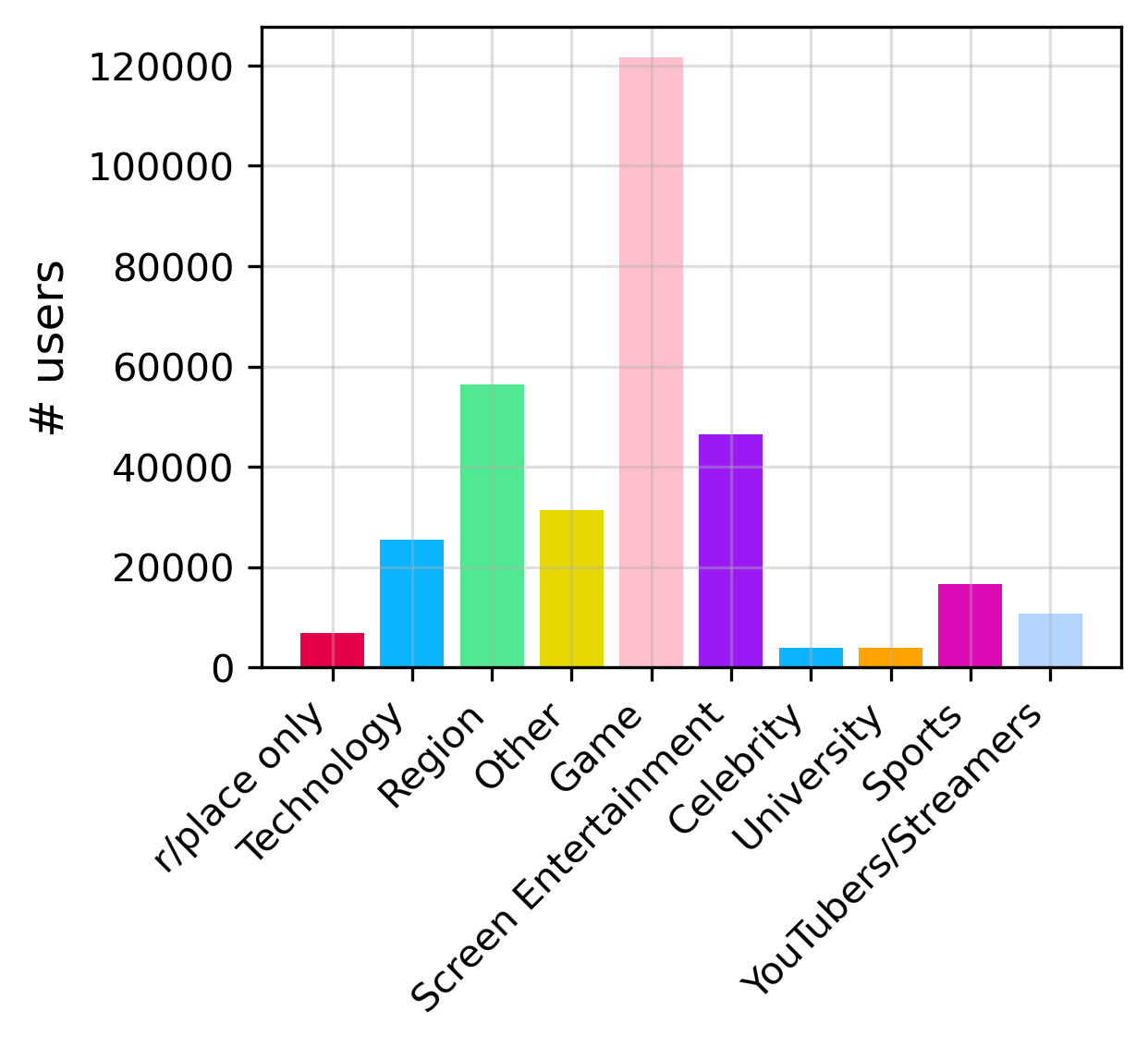}
        \caption{Community population size}
    \end{subfigure}\hfill
    \begin{subfigure}[t]{0.48\textwidth}
        \centering
        \includegraphics[width=0.95\textwidth]{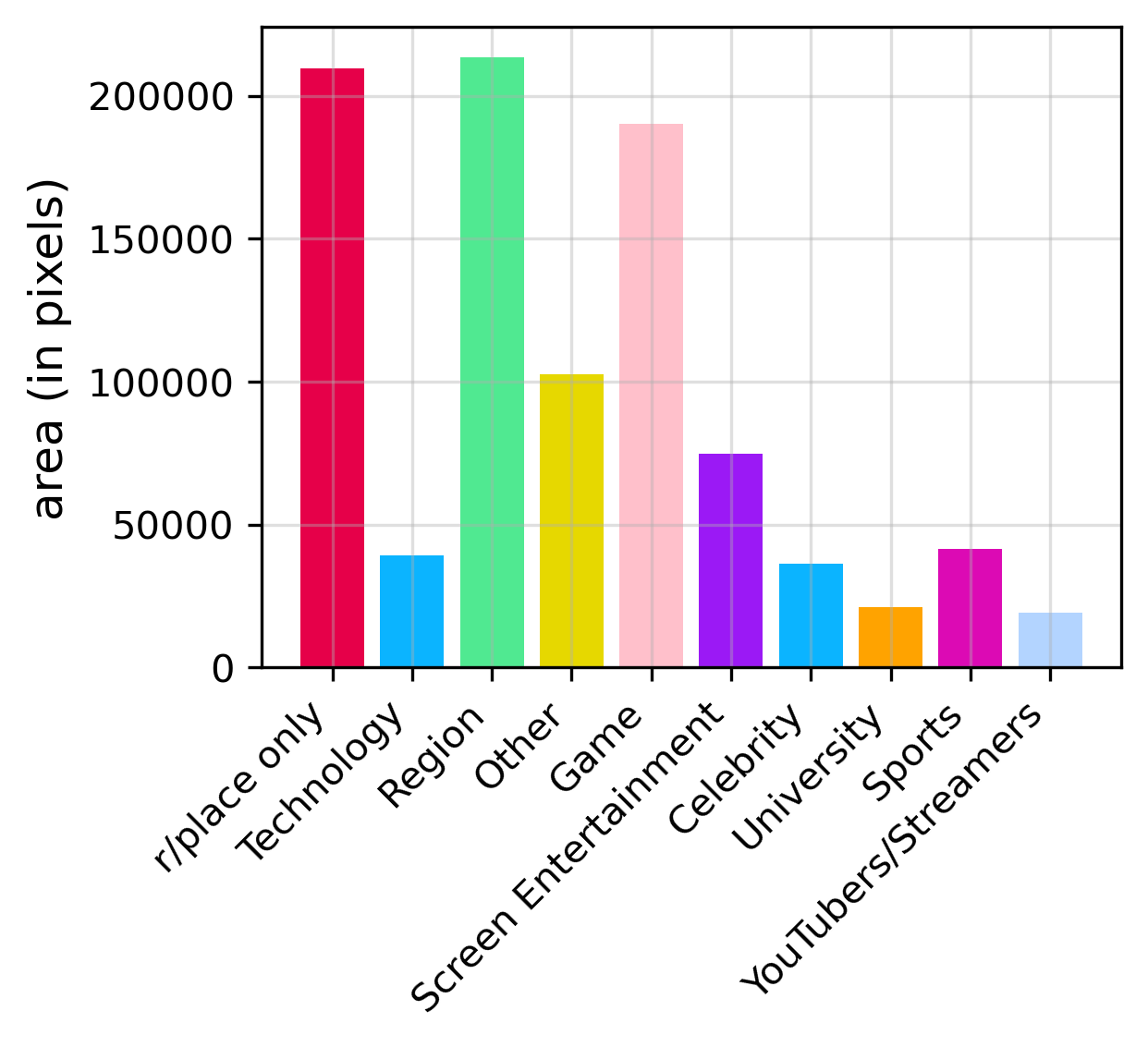}
        \caption{Canvas area per category}
    \end{subfigure}
    \caption{Thematic alignment across subreddit categories during the 2017 event, illustrating the divergence between a community's latent platform scale and its active mobilization capability on the canvas.}
    \label{fig:subreddit_category}
\end{figure}

Figure \ref{fig:subreddit_category} compares subreddit size, measured by active users posting or commenting within a community, against the volume of associated canvas actions. The results reveal a weak correspondence between overall subreddit size and participation in the event. Although large gaming communities dominate subreddits, they did not generate the highest levels of canvas activity. Instead, participation was concentrated among significantly smaller \textit{``r/place only''} communities and geographically oriented groups, including communities centered on countries, regions, and national symbols.

Several highly active coalitions emerged around collective identities rather than conventional topics. For example, communities such as \texttt{r/BlueCorner} coordinated substantial volumes of activity around the maintenance of a shared visual territory despite having a relatively narrow thematic focus. More broadly, the prominence of geographic, cultural, and event-specific communities suggests that participation in r/place was often organized around highly visible collective symbols and shared identities. This interpretation is consistent with prior research showing that group identity can serve as a powerful organizing mechanism in online collective action \cite{tajfel2001integrative,campbell1965ethnocentric,zhang2017community}, particularly in environments that require sustained coordination under resource constraints \cite{kapoor_lee,richerson2016cultural}.

\subsection{Key Takeaways (RQ1)}

Our analysis of engagement dynamics highlights two broader patterns in large-scale peer-production:

\begin{itemize}
\item \textbf{Persistent Participation Inequality Under Rate-Limited Conditions:} Participation exhibits pronounced heavy-tailed and diurnal distributions across all three editions, indicating that activity remains highly concentrated among a relatively small subset of users despite platform-imposed cooldown constraints. These recurring temporal patterns suggest that rate-limited collaborative systems continue to exhibit many of the participation dynamics observed in other large-scale online communities \cite{kwak2010twitter, leskovec2009meme,halfaker2013rise,ortega2008inequality}.

\item \textbf{Participation Is Weakly Related to Community Size:} Canvas participation is only weakly associated with the overall size of a subreddit. The most active coalitions tend to emerge around highly visible collective symbols and shared identities, including geographic, cultural, and event-specific communities. This suggests that participation in large-scale collaborative campaigns may depend less on the size of an existing audience and more on the ability of a community to mobilize around a clear and bounded collective objective.

\end{itemize}

%% file: collaboration.tex
\section{Mass Collaboration (RQ2)}
\label{sec:collaboration}

A coalition is a decentralized collective of participants who coordinate labor toward a shared visual artifact. Operationally, we assign a participant to a coalition if they contribute at least one action whose color matches the final state of the artifact on the closing canvas. Participants are not restricted to a single coalition and may contribute to multiple overlapping efforts throughout the event. As shown in Figure \ref{fig::updates_drawing} and Table \ref{table:data_highlight}, collaboration is highly skewed: the ecosystem contains a small number of very large coalitions and a large number of small, transient ones. To characterize collaborative performance in this competitive environment, we examine how participation patterns and coordination inefficiencies vary across coalitions of different sizes.

\subsection{Coordination Inefficiencies}

Coordination in large-scale collaborative systems requires participants to organize actions, allocate tasks, and integrate distributed labor toward shared goals \cite{cummings_07, Pendharkar_09,kittur2008harnessing}. Prior research suggests that these challenges often become more pronounced as groups grow, increasing the difficulty of maintaining awareness and avoiding redundant work. Because direct measures of communication and planning are unavailable in the r/place logs, we focus on an observable manifestation of coordination inefficiency. Specifically, we define a \textit{wasted action} as an update that repaints a pixel with its current color, producing no visible change to the canvas state. Such actions may arise for several reasons, including imperfect coordination, delayed awareness of recent updates, or redundant defensive behavior. We operationalize coordination inefficiency as the proportion of wasted actions relative to all actions performed by a coalition.

\begin{figure*}[t!]   
    \centering
    \begin{subfigure}[t]{0.33\textwidth}
        \centering
        \includegraphics[width=\linewidth]{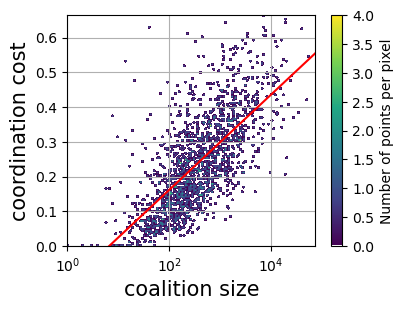}
        \caption{2017}
    \end{subfigure}~
    \begin{subfigure}[t]{0.33\textwidth}
        \centering
        \includegraphics[width=\linewidth]{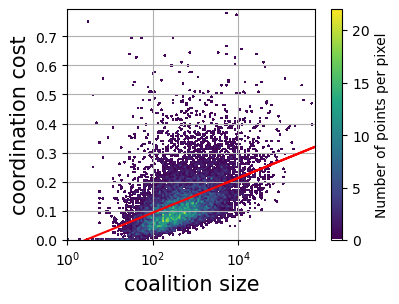}
        \caption{2022}
    \end{subfigure}~
    \begin{subfigure}[t]{0.33\textwidth}
        \centering
        \includegraphics[width=\linewidth]{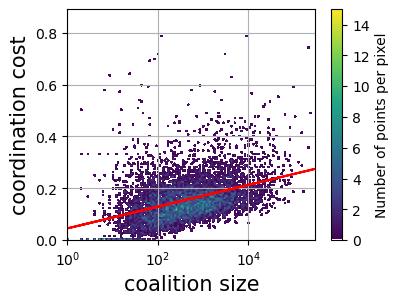}
        \caption{2023}
    \end{subfigure}
    \caption{Longitudinal distribution of coalition size versus coordination inefficiences, operationalized as the proportion of wasted actions over total updates. Results show that inefficiences increase with coalition size but decrease over the three editions of the experiment.}
    \label{fig:coord_cost}
\end{figure*}

Figure \ref{fig:coord_cost} illustrates the relationship between coalition size and coordination inefficiency across the three editions of r/place. In all events, larger coalitions exhibit higher proportions of wasted actions, indicating that redundancy becomes increasingly common as participation scales. However, the strength of this relationship decreased from 2017 to 2023. This cross-edition trend suggests that large coalitions became progressively more effective at limiting redundant work.

Several factors may contribute to this pattern. Over successive editions, communities increasingly relied on external coordination infrastructures, including overlays, browser extensions, bots, and communication platforms. These tools may have helped participants align their actions more effectively and reduce redundancy. At the same time, other factors, including shifts in participant composition, platform conditions, and coalition structure, may also have influenced these trends. Together, the results indicate that while coordination inefficiencies remain associated with coalition scale, their magnitude varies across editions and socio-technical contexts.

\subsection{Participation Patterns Across Coalition Scales}

While technical tools may improve coordination, large-scale peer production also depends on sustained individual participation. Prior work on collective action and group performance suggests that individual effort can vary systematically with group size, including through mechanisms such as social loafing, diffusion of responsibility, and unequal participation \cite{ingham1974ringelmann,kraut2012building,ortega2008inequality,halfaker2013rise}. The highly skewed coalition structure of r/place provides an opportunity to examine how individual contribution patterns vary across organizations of different scales. As shown in Table \ref{table:data_highlight}, the largest coalitions exceed average coalition sizes by factors ranging from 74 to 380$\times$ across the three editions.

\begin{figure*}[t!]   
    \centering
    \begin{subfigure}[t]{0.33\textwidth}
        \centering
        \includegraphics[width=\linewidth]{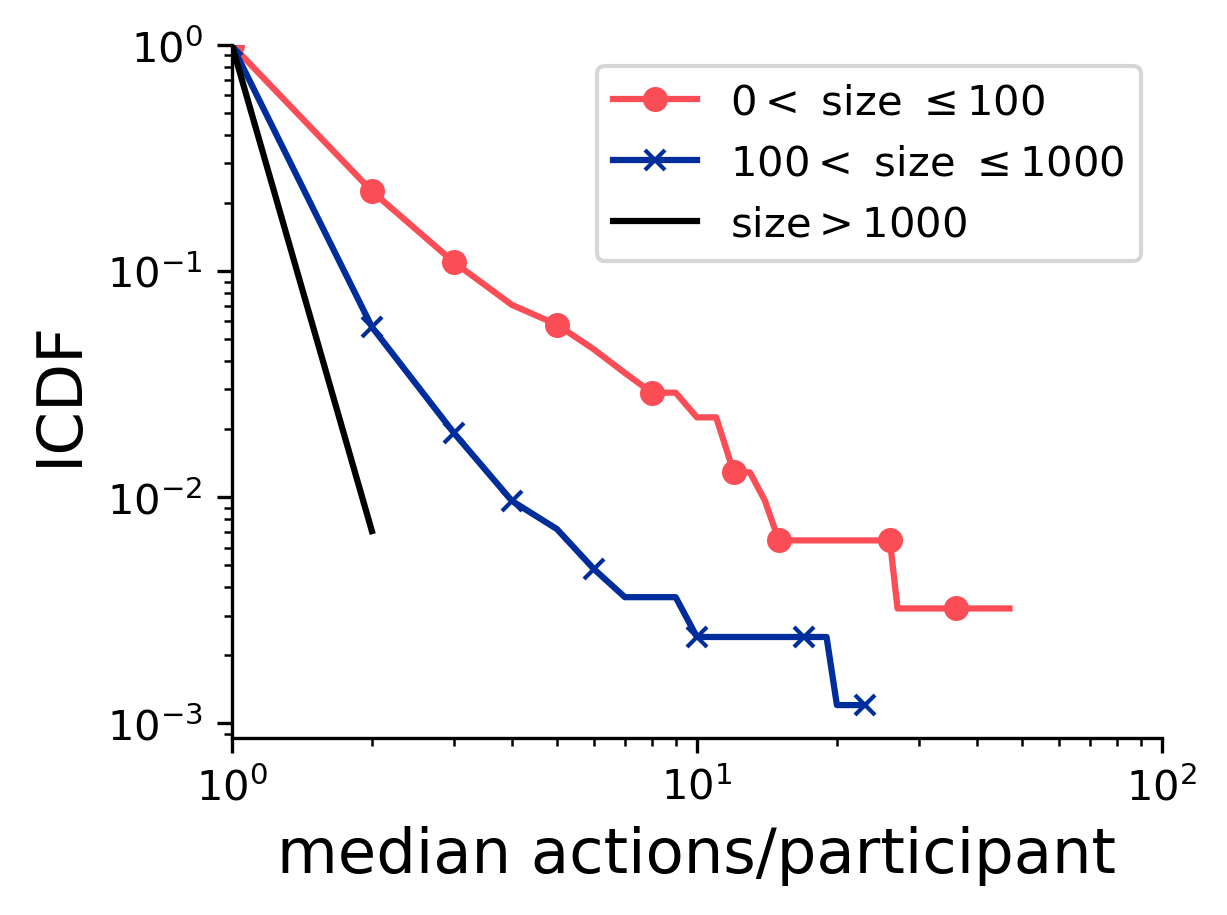}
        \caption{2017}
    \end{subfigure}~
    \begin{subfigure}[t]{0.33\textwidth}
        \centering
        \includegraphics[width=\linewidth]{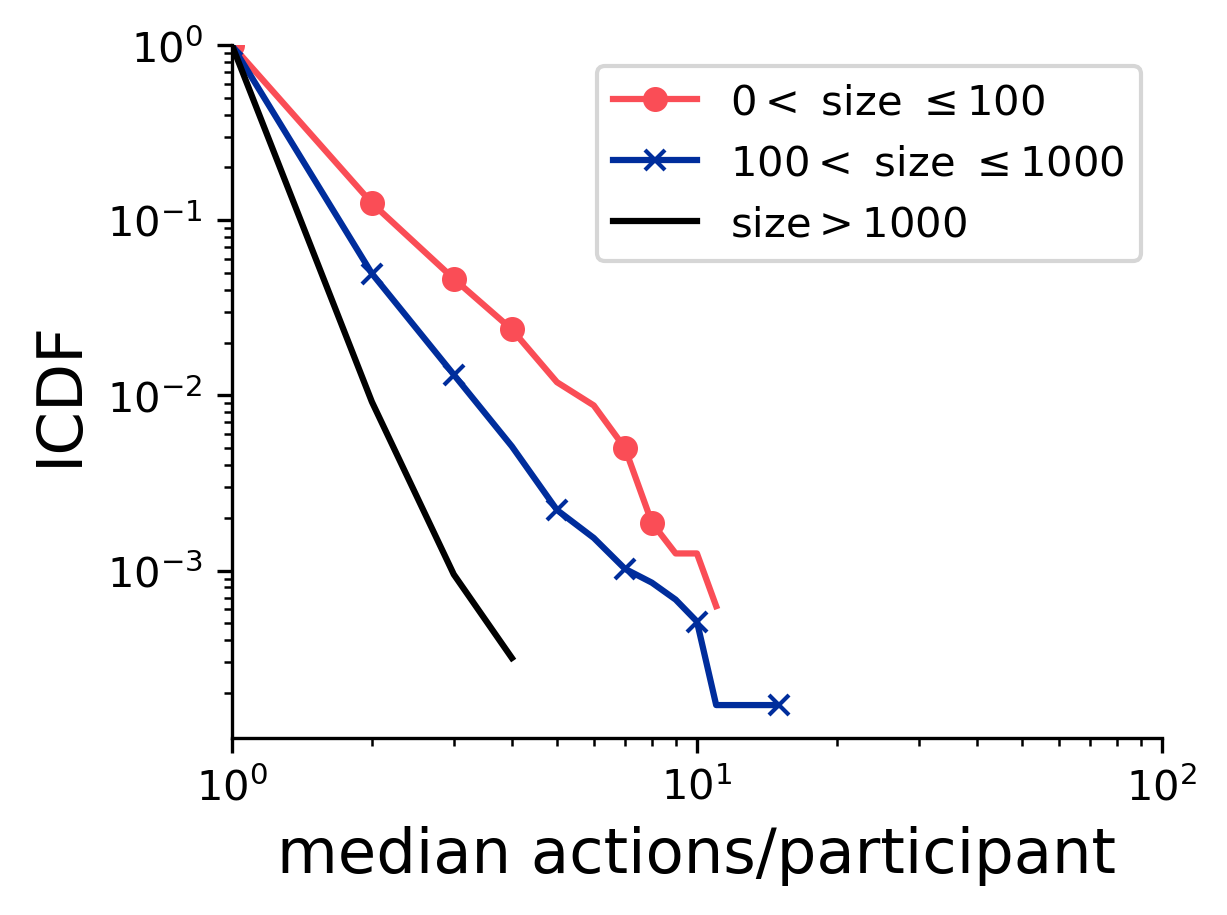}
        \caption{2022}
    \end{subfigure}~
    \begin{subfigure}[t]{0.33\textwidth}
        \centering
        \includegraphics[width=\linewidth]{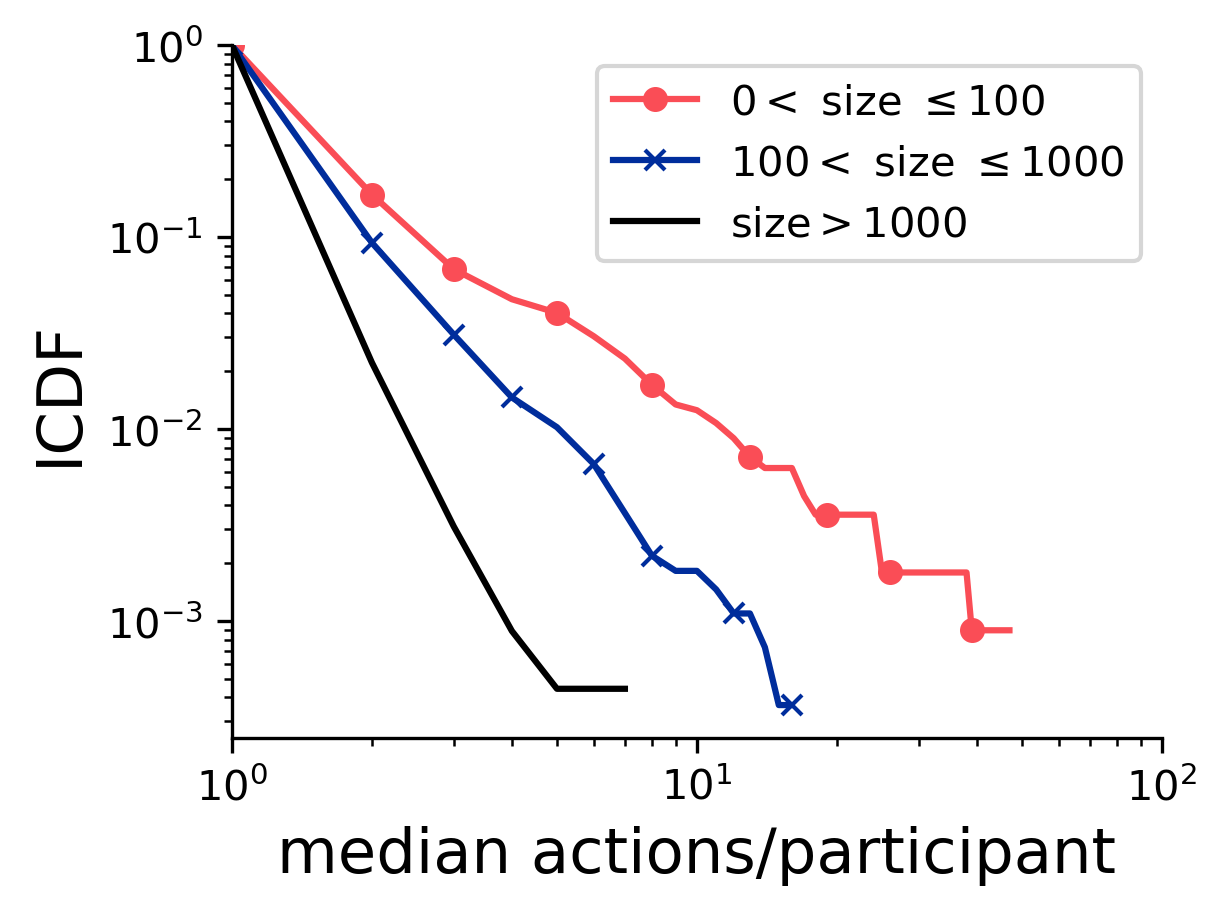}
        \caption{2023}
    \end{subfigure}
    \caption{Cross-edition comparison of individual participation rates stratified by coalition size. Smaller coalitions exhibit substantially higher median per-participant
contribution levels than larger coalitions.}
    \label{fig:ringelmann}
\end{figure*}

Figure \ref{fig:ringelmann} presents the Inverse Cumulative Distribution Function (ICDF) of median actions per participant for three coalition-size categories: small ($\leq100$ participants), intermediate ($100 < \text{size} \leq 1000$), and large ($>1000$ participants). We use coalition medians to reduce the influence of highly skewed participation distributions and extreme contributors within individual groups.

Across all three editions, smaller coalitions exhibit substantially higher median per-participant contribution levels than larger coalitions. This pattern suggests that participation becomes increasingly uneven as coalition size grows, with large groups relying on a broader base of relatively low-activity contributors. The difference is particularly pronounced in the 2017 event, which also exhibits the highest average number of updates per participant (14.19). Several mechanisms may contribute to this relationship, including increased division of labor, the incorporation of peripheral contributors, heterogeneity in member commitment, and other group-size effects identified in the collective-action literature \cite{kittur2008harnessing,kraut2012building}. 



\subsection{Key Takeaways (RQ2)}

Our analysis of collaboration dynamics across the three editions of r/place highlights two broader observations about large-scale peer production systems:

\begin{itemize}
\item \textbf{Coordination Inefficiencies Can Be Moderated by Socio-Technical Infrastructure:} Across all editions, larger coalitions exhibit higher levels of redundant work, suggesting that coordination challenges become more pronounced as groups scale. However, the strength of this relationship decreases over time, indicating that large communities became progressively more effective at limiting redundancy. This trend is consistent with the increasing use of external coordination infrastructures, such as overlays, communication platforms, and automation tools, although additional factors may also contribute.
\item \textbf{Coalition Scale Is Associated with Changing Participation Patterns:} Smaller coalitions consistently exhibit higher median per-participant contribution rates than larger ones. This suggests that participation becomes increasingly uneven as coalition size grows. Several mechanisms may contribute to this relationship, including diffusion of responsibility, division of labor, peripheral participation, and differences in member commitment.
\end{itemize}

These findings reveal a persistent organizational tension within large-scale peer production. As coalitions expand, they face increasing challenges related to coordination and participation. Yet large coalitions remain a recurring feature of the r/place ecosystem. One possible explanation is that collaborative efforts do not emerge in isolation, but instead operate within a competitive environment where groups must continually defend territory and maintain artifact visibility. The next section examines the role of adversarial interactions and conflict in shaping coalition outcomes.

%% file: competition.tex
\begin{figure}[h!]
    \centering
        \includegraphics[width=0.5\textwidth]{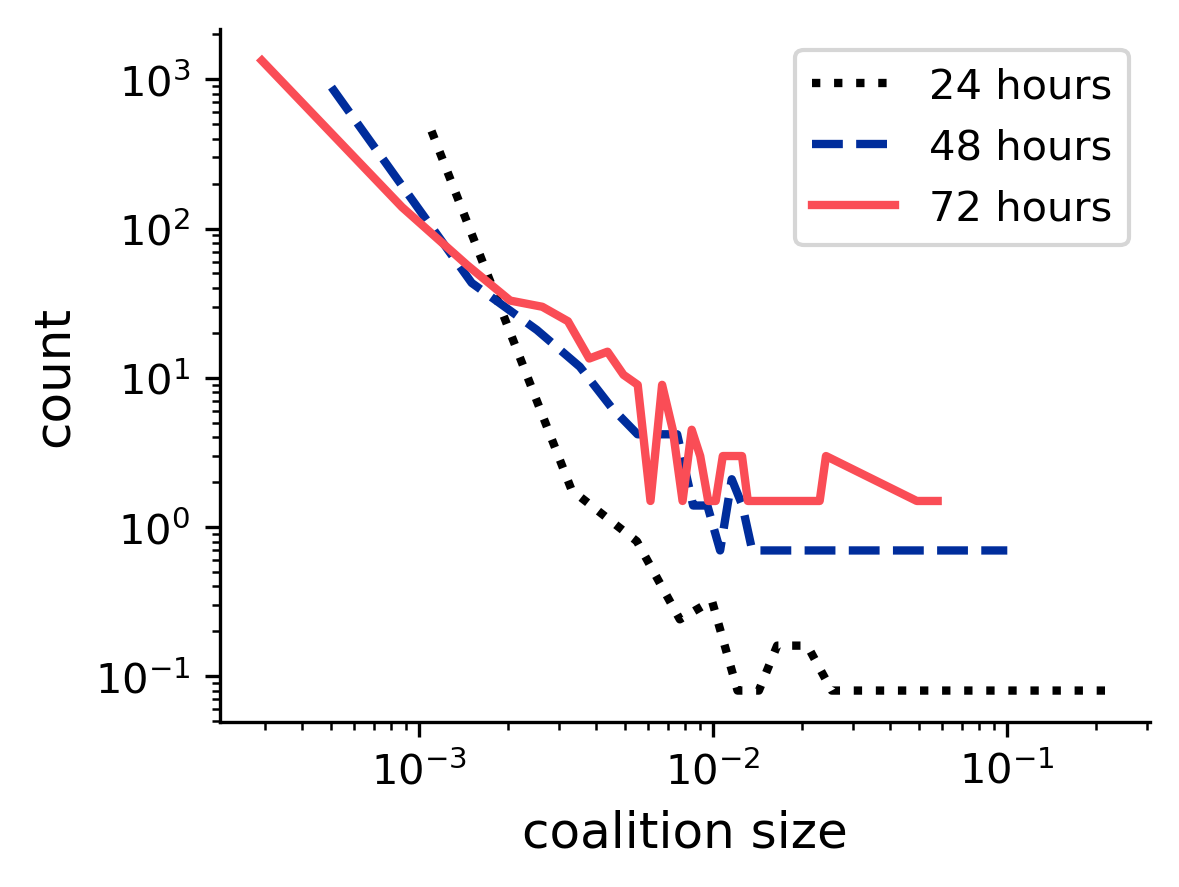}

    \caption{Distributions of coalition sizes relative to the total population for three snapshots of the 2017 edition. Coalition-size distributions become increasingly concentrated in larger groups as the event progresses. 
    }
   \label{fig::power_law}
\end{figure}

\section{Conflict and Outcome Predictability (RQ3)}
\label{sec:conflict}

While platform structures shape patterns of coordination and participation, peer-produced artifacts in r/place also exist within a highly competitive spatial environment. Studying these conflict dynamics presents a methodological challenge due to survivorship bias: community-curated atlases primarily document artifacts that persist until the final canvas snapshot, providing limited visibility into transient or unsuccessful efforts. To address this limitation, we introduce an algorithmic framework that identifies and tracks both persistent and transient coalitions from historical action logs, enabling the analysis of collaborative efforts that are often absent from final-state datasets.

Due to the computational demands of large-scale graph representation learning, we apply our dynamic coalition recovery pipeline to the 2017 dataset. We focus on this edition because it provides a complete record of participant interactions under the core mechanics of r/place. Consequently, the conflict patterns reported in this section should be interpreted as characterizing the 2017 event, while potentially offering insights into broader dynamics that may also be present in later editions.

\subsection{Auditing Transient Peer Production through Coalition Recovery}

To study collaborative efforts that are absent from final-state atlases, we develop a dynamic coalition recovery framework that reconstructs evolving artifacts from historical action logs. Conceptually, the approach generates a sequence of temporally linked canvas atlases, enabling the analysis of both persistent and transient collaborative efforts. Because individual action logs contain only sparse behavioral information ($\langle \text{player}, \text{color}, x, y, \text{time} \rangle$), the framework combines spatial image segmentation with social interaction networks to infer coalition structure. A complete technical description is provided in the Appendix.

\begin{itemize}
\item \textbf{Snapshot Segmentation:} For each canvas snapshot, we apply a scalable graph-based segmentation algorithm to identify spatially contiguous pixel regions with similar color patterns. In parallel, we construct a \textit{player collaboration graph}, where edge weights capture the frequency with which pairs of participants place adjacent pixels of the same color. We then apply \textit{Node2Vec} to learn behavioral embeddings for participants from this interaction network \cite{grover2016node2vec}. Finally, visually localized pixel clusters are merged using Ward's hierarchical clustering based on the similarity of their associated participant embeddings \cite{ward1963hierarchical}.
\item \textbf{Dynamic Coalition Matching:} To connect artifact clusters across consecutive snapshots, we employ a greedy approximation algorithm inspired by the set cover problem \cite{Stergiou-fGreedy}. This procedure links overlapping clusters while remaining computationally tractable at the scale of millions of actions. Historical embedding similarity is then used to refine matches and reconstruct the temporal evolution of coalition activity.

\end{itemize}

\textbf{Evaluation:} We compare the proposed framework against approaches based exclusively on visual information \cite{Felzenswalb-graphcut}, participant interaction \cite{grover2016node2vec}, and the coalition recovery method proposed by \cite{rappaz-place}. Using atlas annotations from the final 2017 canvas as a reference benchmark, our approach achieves an 18\% improvement in Adjusted Rand Score (ARS) and an 8\% reduction in Variation of Information (VI) relative to the strongest baseline. Because no ground-truth annotations exist for transient or failed coalitions, evaluation is necessarily performed against final-state atlas markers. Consequently, this experiment assesses the framework's ability to recover known artifact structure on the final canvas snapshot. We further evaluate the validity of recovered transient coalitions in the Appendix using a combination of ground-truth successful coalitions and visual inspection. 

\subsection{Coalition Consolidation Under Competitive Pressure}

Theories of inter-group competition suggest that competition can encourage greater levels of within-group coordination and organization, particularly when collective outcomes depend on sustained cooperation \cite{richerson2016cultural,lang2022outgroup}. The ability to recover transient and unsuccessful coalitions provides an opportunity to examine how coalition-size distributions evolve over the during the event.

Figure \ref{fig::power_law} shows the distribution of coalition sizes at three milestones (24, 48, and 72 hours), normalized by the active participant population. Across all observations, the distribution shifts toward larger coalitions over time, indicating an increasing concentration of participants within large collaborative groups as the event progresses.

Several factors may contribute to this consolidation pattern. Because the platform restricts users to one action every five minutes, maintaining and defending artifacts requires sustained collective effort. Larger coalitions may therefore possess advantages in coordinating activity, maintaining artifact visibility, and responding to external interference. More broadly, the results suggest that larger organizations may enjoy advantages under sustained competitive pressure.

\subsection{Outcome Predictability}

The competitive nature of r/place raises an important question: to what extent can the eventual success of a coalition be inferred from its early observable characteristics? To explore this question, we evaluate the predictability of coalition outcomes using a set of structural and behavioral features. 

We randomly sample 100 historical canvas snapshots and extract five attributes for each coalition:

\begin{enumerate}
\item \textbf{Start Time:} The normalized timestamp ($[0,1]$) of the coalition's first observed action relative to total event duration.
\item \textbf{Artifact Size:} The area of the artifact measured in pixel coordinates.
\item \textbf{Coalition Size:} The number of unique participants contributing to the artifact.
\item \textbf{Color Entropy:} A measure of visual diversity within the artifact's color composition.
\item \textbf{Success Status:} A binary indicator denoting whether the coalition retained at least 11
\end{enumerate}

The selected features were designed to capture the main observable dimensions of coalition states identified in prior work on collective action and peer production \cite{kittur2008harnessing,kraut2012building}. Coalition size and artifact size capture organizational scale and coordination demands, which have repeatedly been linked to collective performance and coordination costs. Start time captures historical advantages. Finally, color entropy provides a proxy for artifact complexity and the coordination burden associated with maintaining visually heterogeneous designs. Together, these variables span coalition scale, territorial footprint, temporal maturity, historical performance, and artifact complexity—the primary observable characteristics available from behavioral traces alone.

These features yield a dataset of 219,217 observations, comprising 166,696 unsuccessful and 52,521 successful coalitions. After balancing the classes through down-sampling, we train a Decision Tree classifier to evaluate the extent to which coalition outcomes can be predicted from these observable characteristics. Table \ref{tab:class_result} summarizes the classification performance. We have also considered Random Forest, SVM, and Gaussian Processes as classifiers, but they achieve comparable or worse results. 

The results indicate that coalition success is difficult to predict during the early stages of the event using coalition state features alone. For snapshots collected within the first 48 hours, the classifier performs only marginally better than chance. Predictive performance improves later in the event, reaching an F1-score of 0.67 during the final stages. One possible interpretation is that coalition outcomes become increasingly constrained as territorial boundaries stabilize and coalition structures mature. More broadly, these findings suggest that the features considered here provide limited information about long-term coalition success during the early phases of the event, highlighting the complexity and dynamism of large-scale peer production environments.

\begin{table}[b!]
    \centering
    \small
    \begin{tabular}{l|c|c|c}\toprule
    \textbf{Temporal Prediction Window} & \textbf{F1 Score} & \textbf{PR AUC} & \textbf{Baseline Success Rate ($\%$)} \\ \midrule
    Complete Lifecycle (All) & 0.36 & 0.51 & 24.96 \\
    Early Game (0 -- 48 Hours) & 0.09 & 0.39 & 13.02 \\
    Late Game (48 Hours -- End) & 0.67 & 0.60 & 38.12 \\ \bottomrule
    \end{tabular}
    \caption{Predictive modeling of coalition success based on structural features. Low classification accuracy indicates limited outcome predictability from observable coalition-state features.}
    \label{tab:class_result}  
\end{table}


More broadly, the features examined in this analysis primarily describe coalition states rather than coalition interactions. However, coalition outcomes may also depend on interaction-level processes that are difficult to observe from behavioral traces alone, including alliance formation, inter-group negotiations, coordinated attacks, and communication occurring on external platforms. 

\begin{figure}[ht!]
    \centering
    \begin{subfigure}[t]{0.75\textwidth}
        \centering
        \includegraphics[trim={0 0 3cm 0},width=\linewidth]{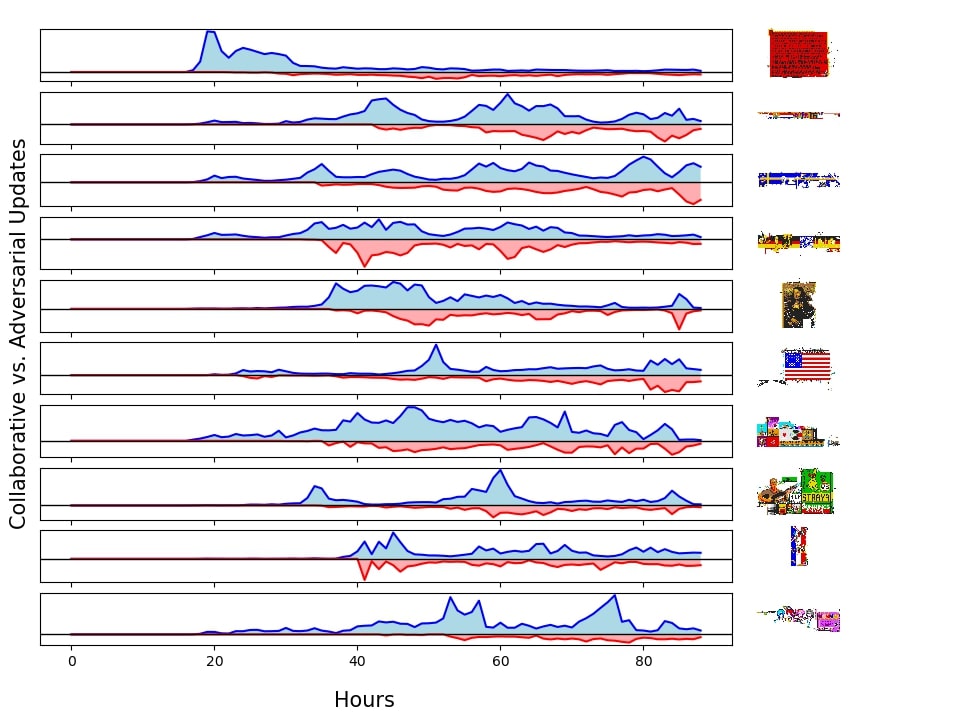}
        \caption{Successful Coalition Lifecycles}
    \end{subfigure}
    
    \begin{subfigure}[t]{0.75\textwidth}
        \centering
        \includegraphics[trim={0 0 3cm 0},width=\linewidth]{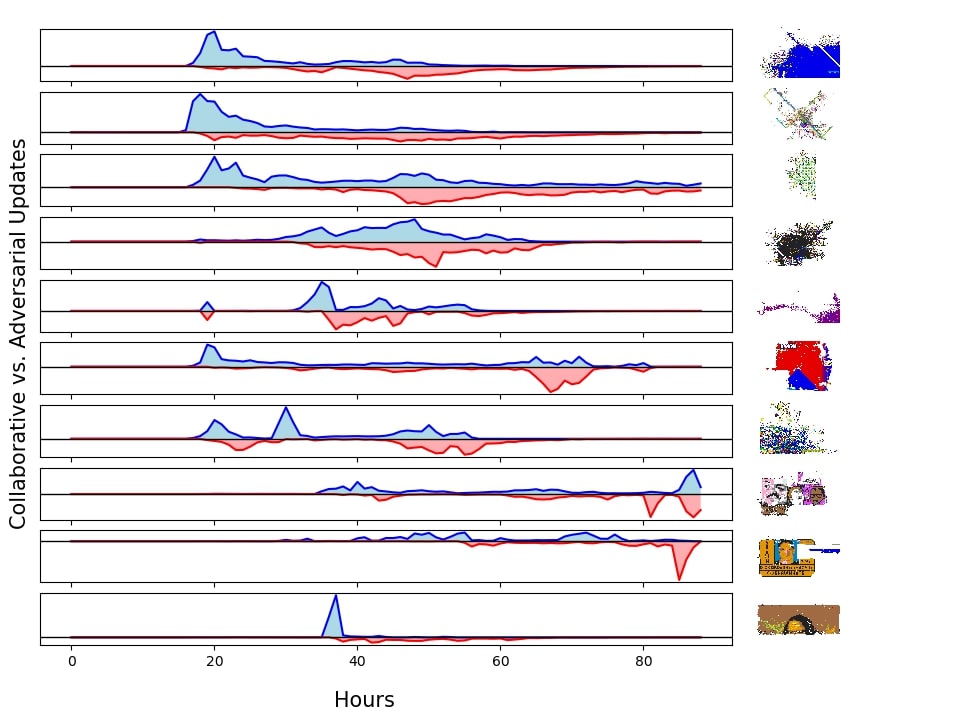}
        \caption{Failed Coalition Lifecycles}
    \end{subfigure}
    \caption{Labor allocation profiles for prominent successful and failed initiatives in the 2017 event. The upper \textcolor{blue}{blue lines} represent collaborative construction actions, while the lower \textcolor{red}{red ones} quantify adversarial overwrites.}
    \label{fig::col_adv}
\end{figure}

\subsection{Successful vs. Failed Coalitions}

To better understand the factors associated with divergent coalition outcomes, Figure \ref{fig::col_adv} traces the temporal labor trajectories of ten representative successful and unsuccessful coalitions, separating collaborative alignment actions from adversarial overwrite activity throughout the event.

The resulting lifecycles reveal substantial variation in how coalitions evolve. Several unsuccessful initiatives (e.g., the \texttt{r/BlueCorner}) accumulated considerable labor during the initial stages of the event yet experienced substantial territorial losses later in the campaign. In contrast, successful initiatives (e.g., \texttt{r/AmericanFlaginPlace}) remained active despite sustained overwrite pressure throughout their lifecycles. Notably, some unsuccessful coalitions exhibited activity patterns similar to successful ones for extended periods before experiencing rapid decline near the end of the event.

These trajectories suggest that early mobilization alone is insufficient to ensure long-term persistence. Instead, coalition outcomes appear to depend on the ability to sustain participation and respond to continued competitive pressure over time. The recovered coalition histories also highlight the importance of repeated mobilization efforts. For example, an initial attempt to construct the \textit{Mona Lisa} was eventually abandoned, whereas a later effort targeting the same artifact achieved sustained territorial presence and reached completion.

Taken together, these lifecycle patterns help contextualize the limited predictability observed in the previous section. Coalitions with similar early activity profiles can ultimately experience very different outcomes, suggesting that artifact persistence is shaped by evolving interactions among participation, coordination, and competitive pressure throughout the event.

\subsection{Key Takeaways (RQ3)}

Our analysis of conflict dynamics highlights two broader implications for the design of large-scale peer-production systems:

\begin{itemize}
\item \textbf{Competition Coincides with Increasing Organizational Concentration:} Coalition growth and persistence occur within a highly contested environment in which artifacts are continually challenged by competing groups. As competitive pressure accumulates, participation becomes increasingly concentrated within larger coalitions, suggesting that sustained collaboration may provide organizational advantages in contested spaces.

\item \textbf{Coalition Outcomes Are Difficult to Predict from Observable Coalition-State Characteristics:} Coalition success remains only weakly predictable during much of the event, even when considering structural characteristics such as coalition size, artifact size, and participation history. This limited predictability suggests that competitive peer-production environments can remain highly dynamic for extended periods, with seemingly similar initiatives ultimately experiencing very different outcomes.

\end{itemize}

These findings suggest that conflict serves not merely as a source of disruption but as a defining feature of the collaborative environment itself. The persistence of large coalitions despite coordination challenges and uneven participation highlights the importance of considering both cooperative and competitive dynamics when designing large-scale collaborative systems.

Because our dynamic coalition lifecycle analysis was limited to 2017, the outcome predictability results reported in Table 2 should be interpreted within the context of a predominantly human-driven collaborative environment, where only 0.3\% of the actions were attributed to accounts filtered as likely automated. Later editions, particularly 2023, exhibit substantially higher levels of high-frequency activity (17.1\%), suggesting a markedly different coordination landscape. 


%% file: discussion.tex
\section{Discussion and Design Implications}
\label{sec:discussion}

We presented a longitudinal, cross-edition analysis of Reddit r/place, examining how engagement, collaboration, and conflict evolved across three large-scale peer-production events. By moving beyond final-state artifacts and leveraging a dynamic coalition-tracking framework, our analysis captures both persistent collaborations and transient initiatives that are absent from community-curated atlases. This broader perspective reveals several recurring organizational patterns. Across editions, larger coalitions exhibited greater coordination inefficiencies and lower median per-participant activity, while competitive pressures were associated with increasing concentration of participants within large collaborative groups. At the same time, coalition outcomes remained difficult to predict during much of the event, underscoring the dynamic and contested nature of large-scale collaborative environments. Taken together, these findings suggest that the organizational dynamics observed in r/place emerge from the interaction between platform constraints, socio-technical infrastructures, and the evolving strategies of participating communities.

\subsection{Synthesizing Cross-Edition Socio-Technical Patterns}

Taken together, the results reveal several recurring patterns in how large-scale peer-production systems operate under conditions of coordination, competition, and resource constraints.

\begin{itemize}
\item \textbf{Participant Engagement (RQ1):} Despite strict rate-limiting constraints, participation consistently exhibits heavy-tailed and diurnal activity patterns across editions. At the same time, engagement is unevenly distributed across communities, with activity concentrating around highly visible and symbolically salient artifacts, including geographic, cultural, and political representations. These findings suggest that participation is shaped not only by platform mechanics but also by the collective meanings attached to shared artifacts.

\item \textbf{Mass Collaboration (RQ2):} Larger coalitions exhibit higher levels of redundant work and lower median per-participant activity, indicating that coordination challenges and participation inequalities become more pronounced as groups scale. However, the relationship between coalition size and redundancy weakens across editions, suggesting that communities became progressively more effective at organizing collective activity. This trend is consistent with the increasing use of external coordination infrastructures and collaborative tools. Nevertheless, scaling collaboration remains associated with organizational challenges that are not fully eliminated by technical support systems.

\item \textbf{Conflict and Outcome Predictability (RQ3):} Competitive interactions are a persistent feature of the r/place ecosystem and coincide with an increasing concentration of participants within large coalitions over time. At the same time, coalition outcomes remain difficult to predict during much of the event, with early observable characteristics providing only limited information about long-term success. Predictability improves later in the event as coalition structures and territorial boundaries become more stable, suggesting that collaborative outcomes emerge through a dynamic process of competition and adaptation.

\end{itemize}

\subsection{Socio-Technical Implications for Collaborative Platform Design}

Beyond characterizing behavior in r/place, our findings highlight several broader design considerations for large-scale collaborative systems operating under conditions of limited resources, decentralized coordination, and inter-group competition. Notably, the emergence of independent pixel-canvas platforms following the original r/place events suggests that these challenges extend beyond a single platform and have motivated a diverse set of design responses. Examining these systems provides a useful lens for considering how collaborative environments might balance participation, coordination, and competition at scale.

\begin{enumerate}

\item \textbf{Supporting Scalable Participation Through Workspace Structuring:} Across all editions, larger coalitions exhibited lower median per-participant activity and greater coordination inefficiencies than smaller groups. These patterns suggest that sustaining meaningful participation becomes increasingly challenging as collaborative efforts scale. One potential design response is to structure large workspaces into smaller, semi-autonomous units that preserve local visibility while maintaining access to a broader collaborative ecosystem. Several pixel-canvas platforms have adopted variations of this approach. For example, \textit{Pixel War: Sub Edition} partitions activity across a collection of isolated 64$\times$64 canvases associated with individual communities, while \textit{Wplace}\footnote{\url{https://wplace.live}} leverages geographic regions as localized participation zones. Although these systems were not evaluated in our study, they illustrate how workspace segmentation can create identifiable subcommunities within a larger collaborative environment and may help participants maintain a clearer connection between individual contributions and collective outcomes.

\item \textbf{Balancing Competition and Outcome Stability:} Our analysis indicates that competitive interactions are a defining feature of the collaborative environment, yet coalition outcomes remain difficult to predict during much of the event. While competition can encourage participation and strategic coordination, highly volatile environments may also make long-term collaborative investments difficult to sustain. Platform designers may therefore benefit from exploring mechanisms that balance openness and contestation with some degree of outcome stability. Contemporary canvas platforms have experimented with a variety of approaches. For instance, \textit{Pxls.space}\footnote{\url{https://pxls.space/}} employs adaptive cooldown policies that vary according to platform activity levels, while \textit{Wplace} incorporates territorial defense mechanics that provide limited advantages to local communities. These examples highlight different ways of moderating competitive pressure without eliminating conflict entirely. More broadly, they suggest that collaborative systems can shape the intensity and consequences of competition through platform-level design choices.

\item \textbf{Integrating Coordination Infrastructure Into the Platform:} The weakening relationship between coalition size and redundant work across editions is consistent with the increasing use of external coordination tools. While these tools may help communities align actions and manage large-scale collaboration, they can also create disparities between groups with different levels of technical sophistication. One response is to incorporate coordination support directly into the platform interface rather than relying exclusively on external overlays, communication channels, or custom scripts. Platforms such as \textit{Pxls.space} demonstrate the feasibility of this approach through built-in activity visualizations, collaborative templates, and planning tools. By embedding coordination infrastructure within the platform itself, designers can reduce barriers to participation and make organizational resources more accessible. Such features may help narrow the gap between highly organized groups and participants who lack access to specialized technical infrastructure.

\end{enumerate}

Taken together, these examples underscore a broader design challenge: collaborative systems must simultaneously support participation, coordination, and competition. The evolution of r/place and its successors suggests that these objectives are deeply interconnected, and that platform design choices can meaningfully influence how communities organize, coordinate, and compete within shared digital environments.

\subsection{Limitations and Future Research Directions}

Several limitations should be considered when interpreting our findings. First, the privacy-preserving anonymization of the released datasets prevents linking participant identifiers to broader platform histories or demographic information. Consequently, our analysis focuses on observable behavioral traces and cannot directly examine how prior community membership, user experience, or communication practices influence participation and coalition outcomes. We mitigate this limitation by leveraging aggregate cross-platform text traces from the Reddit API.

Second, our dynamic coalition recovery framework reconstructs coalition trajectories from behavioral logs rather than directly observing ground-truth structures. Although the approach performs well against available benchmarks, the validity of recovered transient and unsuccessful coalitions cannot be verified directly because no comprehensive annotations exist for these artifacts.

Third, the computational demands of large-scale graph representation learning constrained our detailed coalition lifecycle analysis to the 2017 edition. While the core mechanics of r/place remained consistent across editions, the conflict dynamics presented in this study should therefore be interpreted primarily as characterizing the 2017 event. Extending coalition recovery to the substantially larger 2022 and 2023 datasets remains an important direction for future work.

Future research could build on this foundation by integrating behavioral traces with communication data from platforms such as Reddit and Discord. Combining spatial actions with textual coordination records would enable a richer understanding of how communities organize collective action, negotiate conflict, allocate tasks, and adapt to changing platform conditions. More broadly, such multimodal analyses could help clarify the relationship between communication, coordination, and organizational outcomes in large-scale peer-production systems.

%% file: appendix.tex
\section{Appendix}

\subsection{Code and data availability}
All implementations used in our experiments in \url{https://anonymous.4open.science/r/r-place-B854/}

Datasets for the three editions of r/place:
\begin{itemize}
\item \url{https://console.cloud.google.com/storage/browser/place_data_share} (2017)
\item \url{https://placedata.reddit.com/data/canvas-history/index.html} (2022)
\item \url{https://placedata.reddit.com/data/canvas-history/2023/index.html} (2023)
\end{itemize}

\subsection{Subreddit categories}
Table \ref{table:a_category_explanation} describes the categories in Figure \ref{fig:subreddit_category}.

The prompt to generate the categorizations is as follows: 

\begin{mdframed}[backgroundcolor=gray!15, roundcorner=4pt, linecolor=gray!40]
\textit{``Can you help me categorize the following subreddits? Here are the categories: YouTubers/Streamers, Technology, r/place only, Games, Screen Entertainment, Sports, Celebrity, University, and Regions. For the category r/place only, it means this subreddit is created to discuss the r/place experiment. For regions, it can be cities or countries. When you are done, please return a spreadsheet with 3 columns. The first column is the given subreddit name, the second column is the category, and the third column is the description.''}
\end{mdframed}

\begin{table*}[ht!]
\centering
    \begin{tabular}{ |c|m{20em}|c|} 
    \hline
     \textbf{Category Name} & \textbf{Description} & \textbf{Example}\\
     \hline
     r/place only & Created specifically to discuss the r/place event. & r/BlueCorner\\
     \hline
     Technology & Tech-related discussions. & r/linux \\
     \hline
     Region & Discussions related to regions that can be observed on a map, such as nations or cities. & r/Switzerland\\
     \hline
     Game & Discussions related to video games, or specific characters of a video game. & r/osugame\\ 
     \hline
     Screen Entertainment & Screen media discussions, such as TV shows, films, animes, etc. & r/gameofthrones\\ 
     \hline
     Celebrity & Celebrities including singers, actors, etc. & r/TheBeatles\\ 
     \hline
     University & Subreddits for colleges \& universities. & r/UTAustin\\ 
     \hline
     Sports & Subreddits for discussing sports events or sports teams. E-sports teams are also included.
     & r/SydneyFC
     \\
     \hline
     YouTubers/Streamers & Subreddits to discuss YouTubers or Streamers. & r/loltyler1\\
     \hline     
     Other & Other subreddits not in the categories listed above. & r/SCP\\
     \hline
    \end{tabular}
    \caption{Description of subreddit categories.}
    \label{table:a_category_explanation}
\end{table*}

\subsection{Coalition Recovery Algorithm}
\subsubsection{Snapshot segmentation}
The first part of our algorithm segments pixels in each snapshot of the canvas into clusters. We describe three segmentation approaches (two existing and our new solution). These approaches differ by the type of features they are able to leverage (visual, social, or both) and should be accurate and scalable.

\textbf{Graph-based Image Segmentation (GBIS) \cite{Felzenswalb-graphcut}:}
We segment the actions in each snapshot $U_t$ based on RGB values by adapting a graph-based image segmentation method \cite{Felzenswalb-graphcut} to our scenario. First, we construct a graph $G^{rgb}_t = (V^{rgb}_t, E^{rgb}_t)$ for each snapshot, where $v \in V^{rgb}_t$ denotes an action  and $e_{uv} \in E^{rgb}_t$ if and only if actions $u$ and $v$ are adjacent. We calculate the weight of edges as $w_{uv}=||\mathbf{c}_u-\mathbf{c}_v||_2$, where $\mathbf{c}_u$ is the RGB code for the color of action $u$. The resulting graph $G^{rgb}_t$ is segmented using the algorithm proposed in \cite{Felzenswalb-graphcut}, which is a recursive algorithm based on a parameter $\kappa$. Initially, each action is assigned to its own cluster. Clusters are merged based on comparisons between edge weights within and across clusters, and $\kappa$ allows such criteria to be adaptive to cluster sizes---i.e., more strict as clusters grow. A key advantage of this algorithm is that it runs in time $O(n\log(n))$, where $n=|V^{rgb}_t|\leq 1000^2$.

\textbf{Ward agglomerate clustering of participant embeddings (N2V-Ward) \cite{ward1963hierarchical,grover2016node2vec}:} While GBIS only captures color information within a snapshot, here we focus on capturing participant activity via vector representations. Similar to the approach from \cite{rappaz-place}, pixels are then clustered based on the representations of their authors using a hierarchical clustering algorithm \cite{ward1963hierarchical}. We propose applying Node2Vec (N2V) embeddings  \cite{grover2016node2vec} to a participant activity graph $G_{participant} = (V_{participant}, E_{participant})$, where nodes $v \in V_{participant}$ are participants and $e_{uv} \in E_{participant}$ if and only if two participants made actions at neighboring pixels with the same color. Intuitively, these connections are likely to represent collaborations. As a result of the embedding algorithm, we obtain a matrix $M^{participant} \in \mathbb{R}^{N\times h}$, where $N$ is the number of participants and $h$ is a parameter determining the number of dimensions of the embeddings. After assigning the vector $M^{participant}[u.participant]$ to each action $u$ in the snapshot $U_t$, we cluster spatially adjacent actions in the snapshot using the Ward algorithm \cite{ward1963hierarchical}. At each iteration, the algorithm groups actions that minimize the total within-cluster variance (similar to K-means) while a threshold $\delta$ is satisfied. The running time of the Ward is $O(n^2)$, where $n$ is the number of actions in a snapshot.

\begin{algorithm}
    \caption{GBIS-N2V-Ward Algorithm for Activity Segmentation}
    \label{alg:segmentation}
   
    \begin{algorithmic}[1]
    \Require Actions $U_t$ for snapshot $t$, N2V participant embedding matrix $M^{participant}$  
    \Ensure Clusters $C_t$ for pixels in the snapshot
        \State $G_t^{rgb} \Leftarrow \text{Grid graph for pixels in } U_t$
        \State $C_t^{rgb} \Leftarrow \text{GBIS segmentation of } G_t^{rgb}$
        \For{each cluster $c$ in $C_t^{rgb}$}
        \State $M^{cluster}[c] \Leftarrow \frac{1}{|c|} \sum_{u \in c} M^{participant}[u.participant]$
        \EndFor
        \State $C_t \Leftarrow \text{Ward clustering of }M^{cluster}$ 
    \end{algorithmic}  
\end{algorithm}

\textbf{Combining participant embeddings and image segmentation (GBIS-N2V-Ward):} We combine these different sources of information using Algorithm \ref{alg:segmentation}. It receives as parameters the set of actions $U_t$ in a given snapshot and the  N2V participant embedding matrix $M^{participant}$ based on the participant activity graph $G_{participant}$. In the first phase, it applies the GBIS segmentation to segment the actions as $C^{rgb}_t$ (lines 1-2). Next, it treats each cluster identified in the first phase as a \textit{super-action} (similar to \textit{super-pixels} in computer vision) and assigns the average participant embedding $M^{cluster}[c]$ of the authors of the super-action $c$ (lines 3-4). Finally, Ward clustering is applied to the super-actions based on their embeddings (line 5).

\subsubsection{Experimental results}
\label{sec::segmentation_results}

\begin{figure*}
    \centering
    \subfloat[Ground-truth]{{\includegraphics[width=0.24\textwidth]{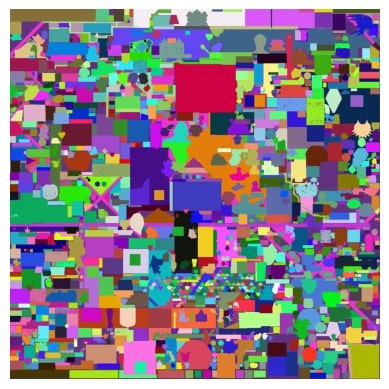}}}
        \subfloat[GBIS]{{\includegraphics[width=.24\textwidth]{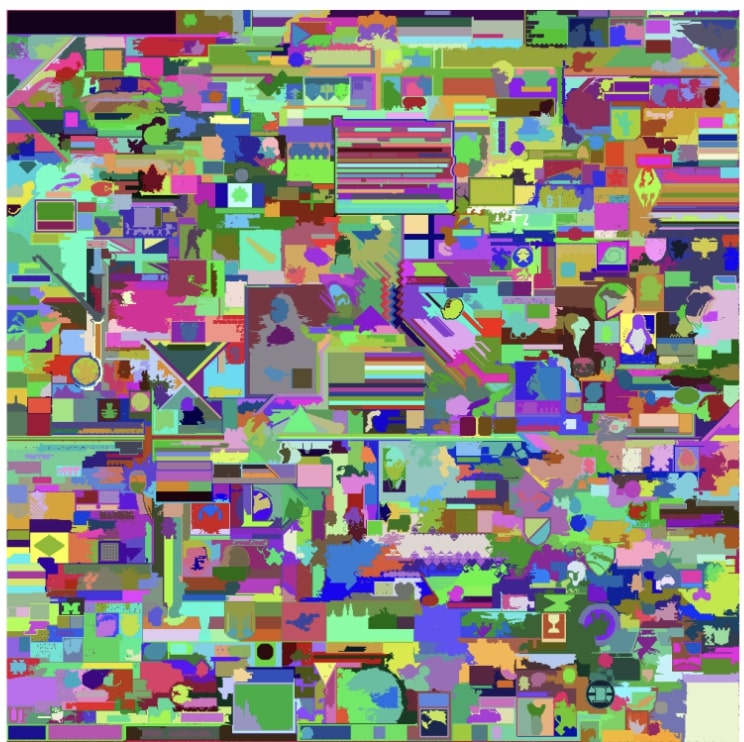}}}
        \subfloat[BPR-Ward]{{\includegraphics[width=.24\textwidth]{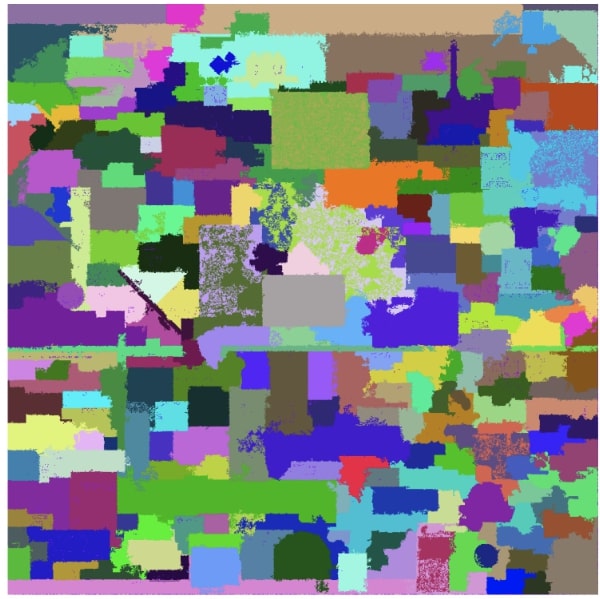}}}
        \subfloat[GBIS-N2V-Ward]{{\includegraphics[width=.24\textwidth]{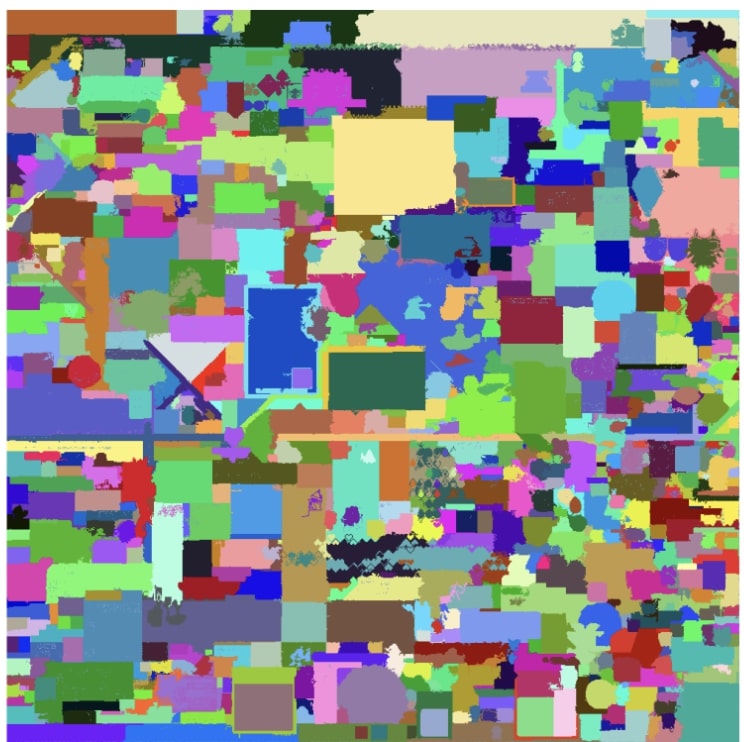}}}
\caption{Comparison of segmentations on the final canvas using three techniques. 
}
\label{seg_picture}
\end{figure*}


We use 10 snapshots of the 2017 dataset to represent the beginning, middle, and end of the experiment.

\textbf{Baselines:} We compare our method against the following: \emph{GBIS} is a graph-based method that accounts only for visual features; \emph{N2V-Ward} combines Node2Vec embeddings of the participant activity graph and Ward clustering; \emph{BPR-Ward} combines BPR embeddings and Ward clustering, as proposed in \cite{rappaz-place}; \emph{GBIS-BPR-Ward} is similar to our approach but applies the BPR embeddings instead of Node2Vec.

\textbf{Accuracy metrics:} We apply \emph{Adjusted Rand Score (ARI)} and \emph{Variation of Information (VI)} \cite{meilua2003comparing,hubert1985comparing}:

\begin{align}
    ARI = \frac{RI(X,Y)- \text{Expected Index}(X,Y)}{max(RI) - \text{Expected Index}(X,Y)}
\end{align}
where $RI$ is the rand index calculated as
\begin{align}
    RI = \frac{TP + TN}{TP + FP + FN + TN}
\end{align}
\begin{align}
    VI(X,Y) = H(X) + H(Y) - 2I(X, Y)
\end{align}
where $H$ is entropy, and $I$ is mutual information. As these metrics require ground-truth labels, we mask the actions that are not tagged in the atlas. We also show the running time to illustrate some of the scalability challenges in handling our datasets.

\begin{table}
    \centering
\begin{tabular}{c|c|c|c}\toprule
{Methods}   &  {ARS ($\uparrow$)} & {VI ($\downarrow$)}& {Runtime (sec)}\\  \midrule
GBIS 
&0.26 & 2.34 & 79\\ 
N2V-Ward & 0.49& 1.50 & 2735\\ 
BPR-Ward & 0.21 & 3.45 & 2646\\ 
GBIS-BPR-Ward & 0.25 & 3.90 & 974\\ 
\textbf{GBIS-N2V-Ward}  &  \textbf{0.58} & \textbf{1.38} & 847\\ 
\bottomrule
\end{tabular}
\caption{Segmentation accuracy in terms of ARS and VI and running time (in seconds). 
}
    \label{tab:experiment_results}
\end{table}

\textbf{Results:} Table \ref{tab:experiment_results} and Figure \ref{seg_picture} show the evaluation and the segmentation produced by the different approaches, respectively. The poor GBIS results indicate that visual features alone are not sufficient for identifying the digital artifacts---it either breaks down individual artifacts based on color or merges adjacent artifacts depending on the value of $\kappa$. While N2V-Ward achieves better results, it still fails to identify sharp borders between certain artifacts, which are key for distinguishing collaboration and competition during the experiment. BPR-Ward suffers from similar issues as N2V-Ward but achieves even worse results than GBIS. The main advantages of the N2V embeddings compared to the BPR ones are that they exploit some color information---only actions with the same color produce an edge---and they account for connectivity beyond one-hop via random walks. Our approach (GBIS-N2V-Ward) outperforms all the alternatives in terms of both accuracy metrics and is an order of magnitude faster than N2V-Ward. However, notice that GBIS-BPR-Ward, which combines GBIS and BPR-Ward, does not achieve comparable results with our approach, which we also believe is due to limitations of BPR embeddings.

 \begin{figure*}[h!]
    \centering
    \begin{subfigure}[t]{0.32\textwidth}
        \centering
        \includegraphics[width=\linewidth]{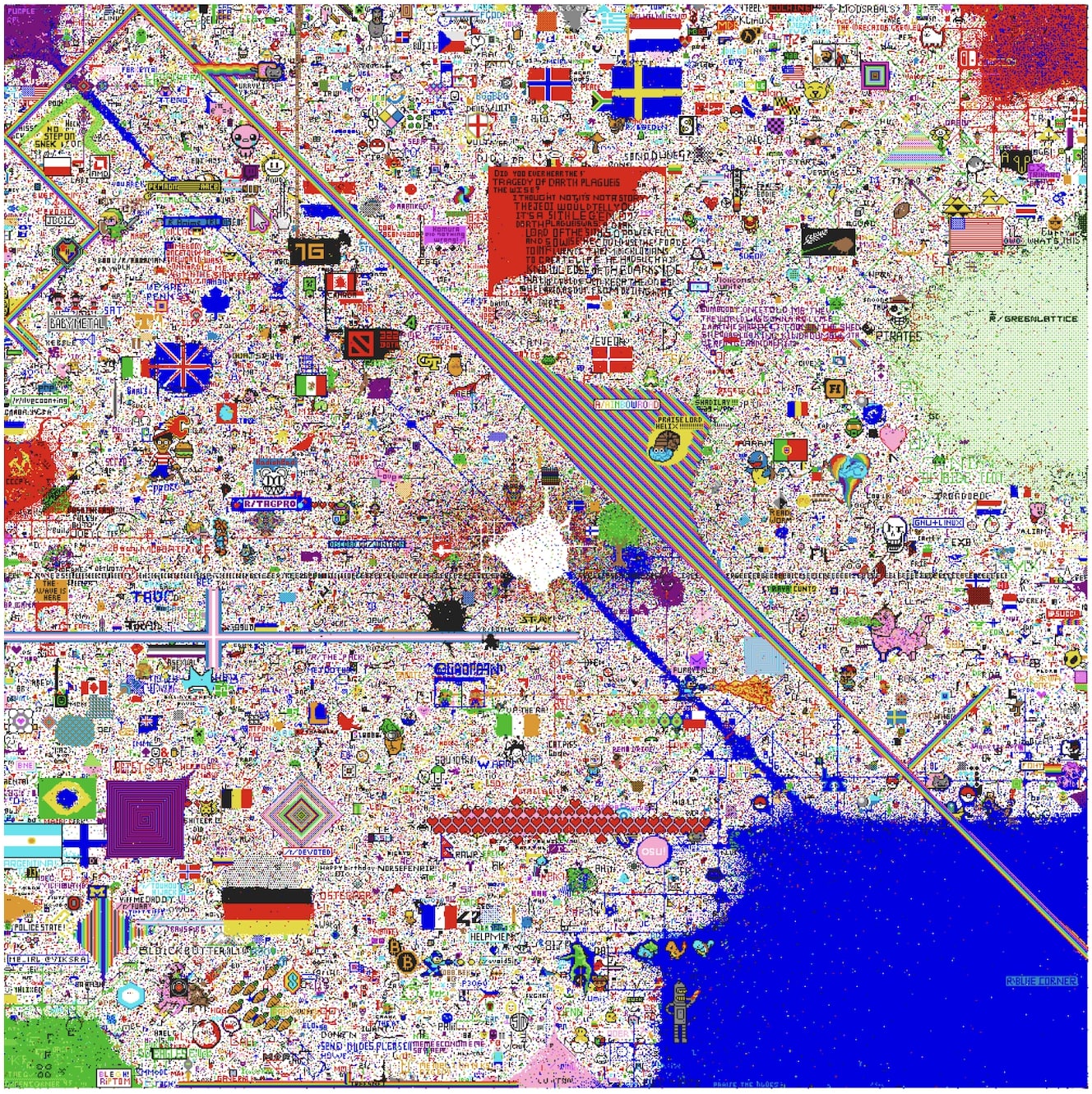}
        \caption{Canvas at hour 27}
    \end{subfigure}~        
    \begin{subfigure}[t]{0.32\textwidth}
        \centering
        \includegraphics[width=\linewidth]{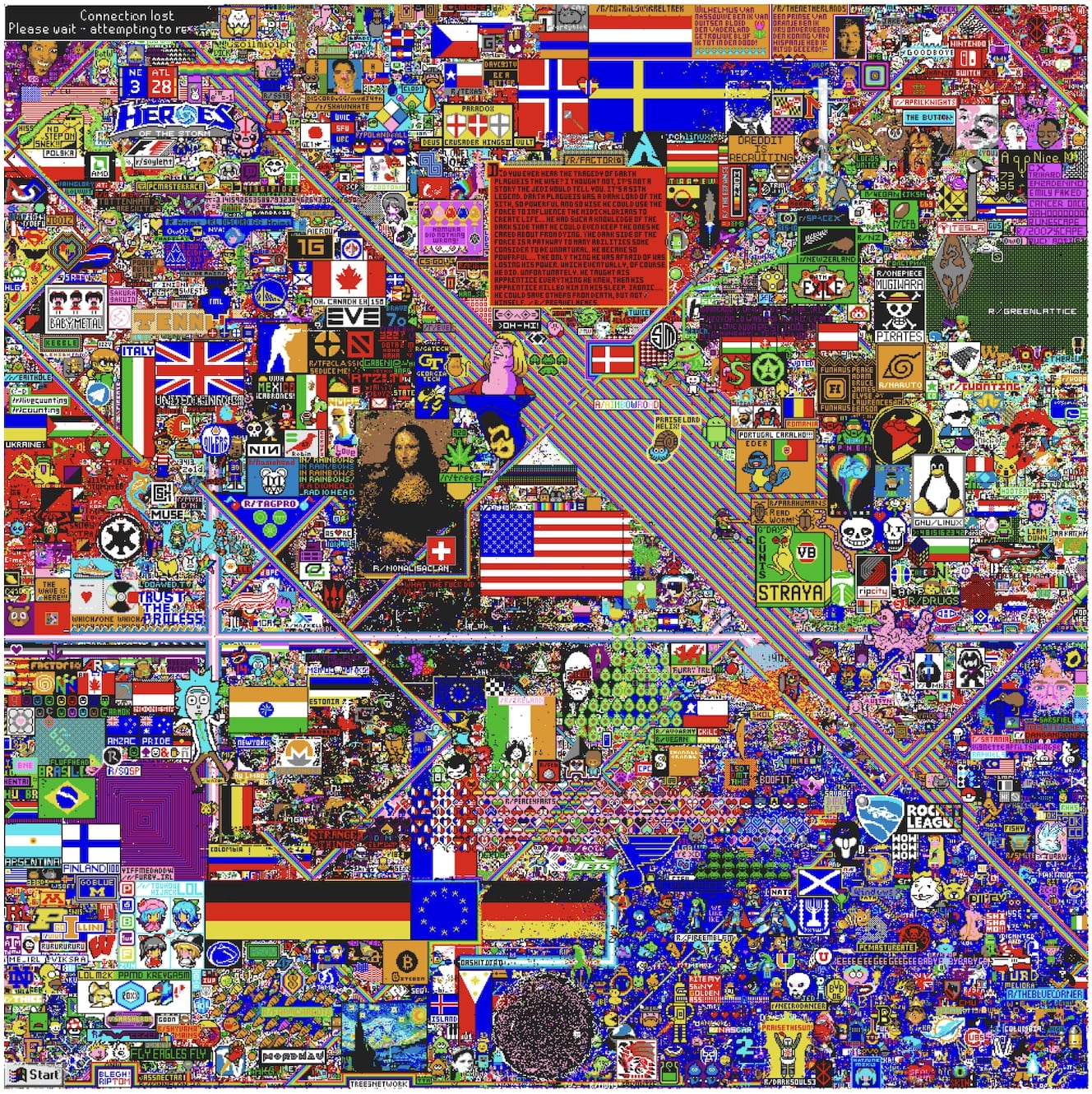}
        \caption{Canvas at hour 55}
    \end{subfigure}~
        \begin{subfigure}[t]{0.32\textwidth}
        \centering
        \includegraphics[width=\linewidth]{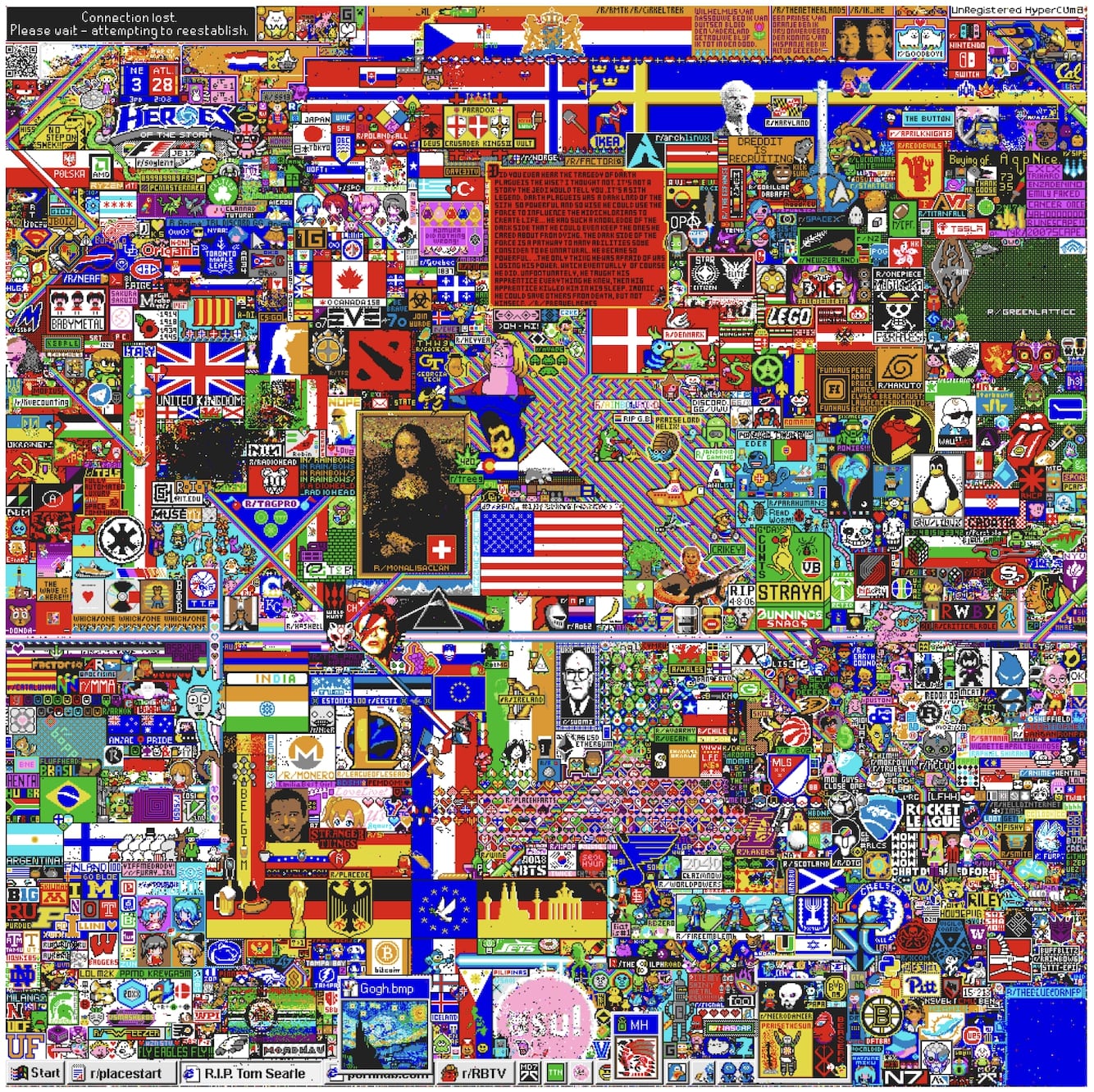}
        \caption{Final canvas}
    \end{subfigure}
    
        \begin{subfigure}[t]{0.32\textwidth}
        \centering
        \includegraphics[width=\linewidth]{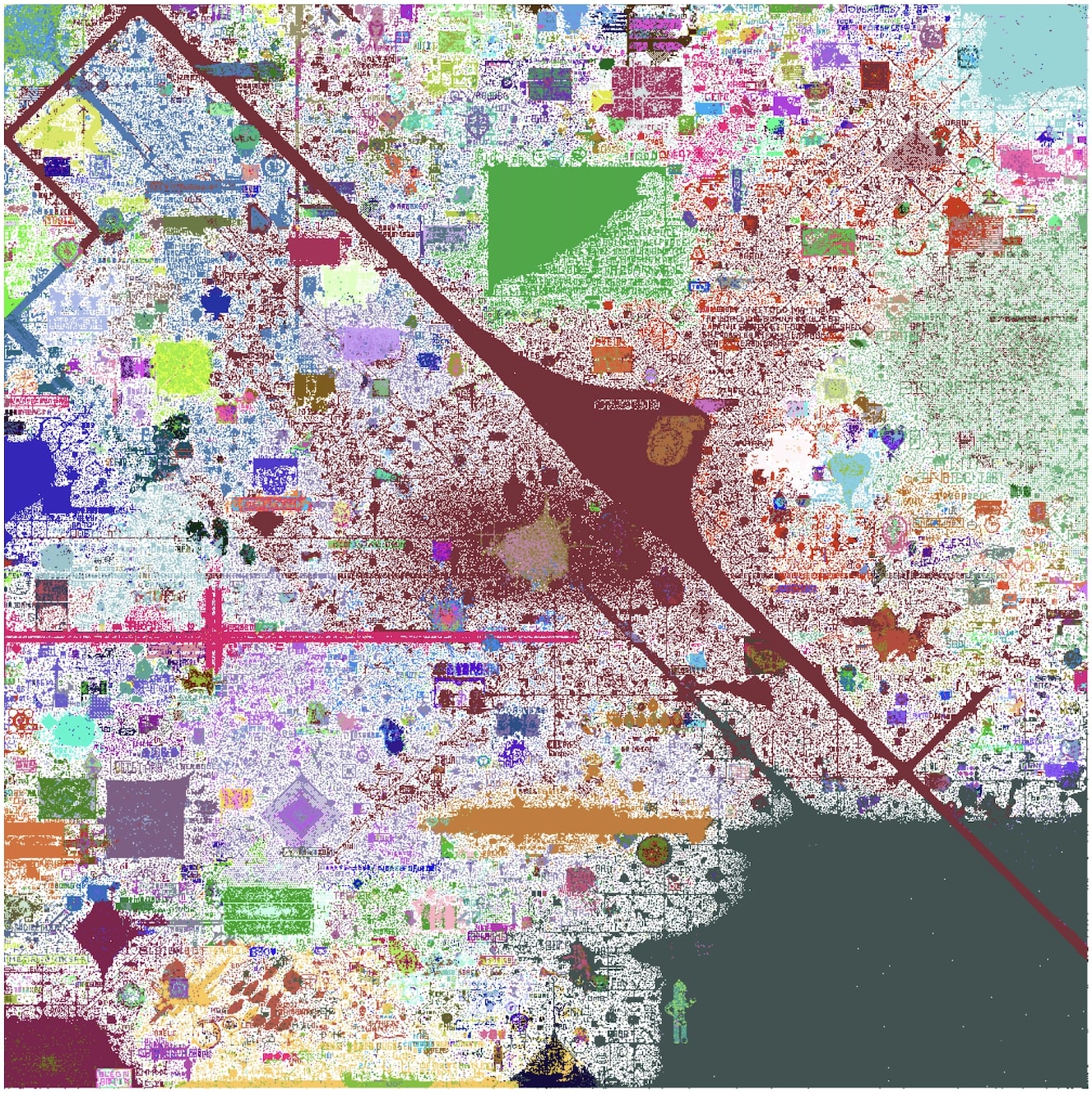}
        \caption{Clustering at hour 27}
    \end{subfigure}~
    \begin{subfigure}[t]{0.32\textwidth}
        \centering
        \includegraphics[width=\linewidth]{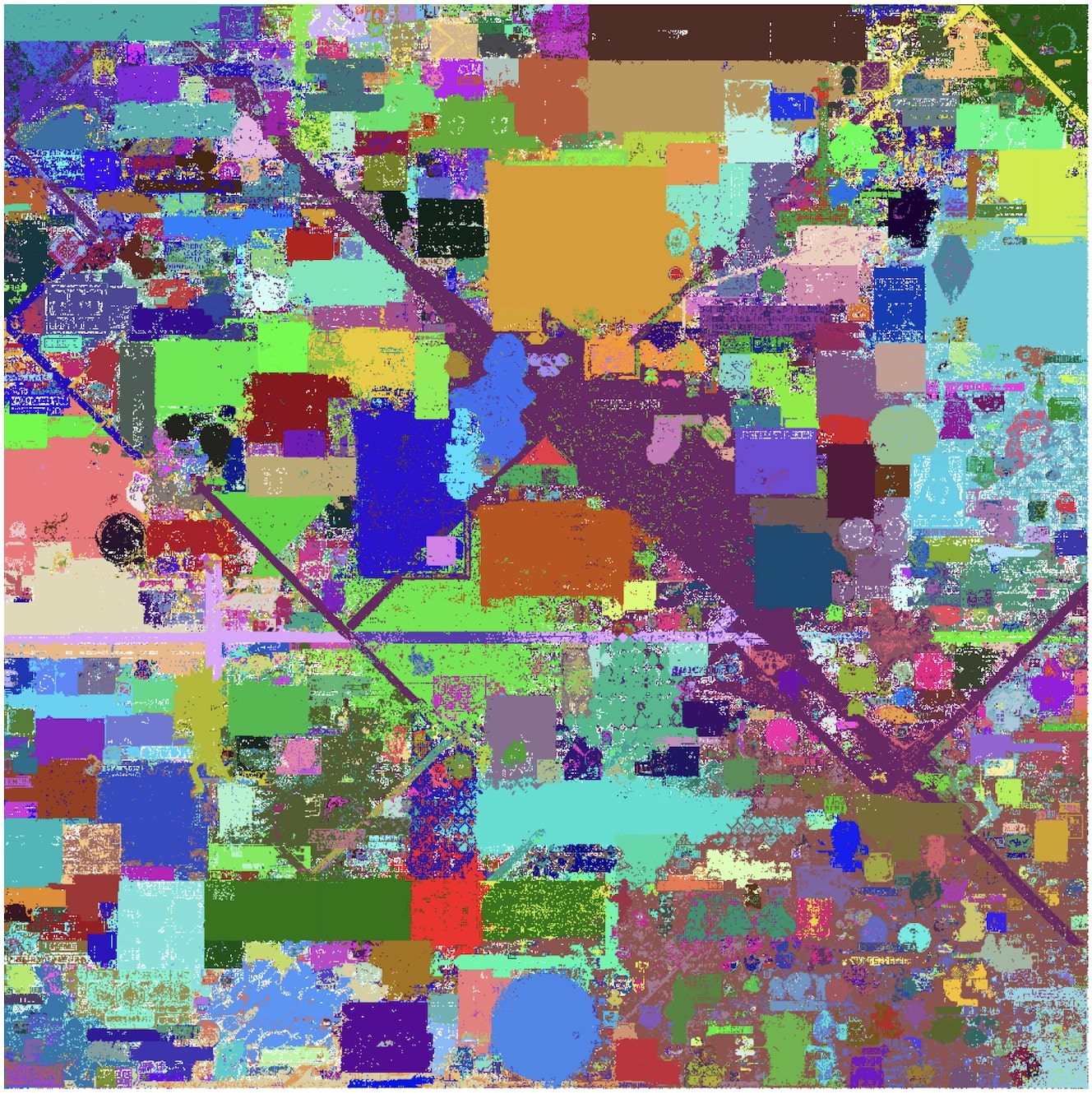}
        \caption{Clustering at hour 55}
    \end{subfigure}~
            \begin{subfigure}[t]{0.32\textwidth}
        \centering
        \includegraphics[width=\linewidth]{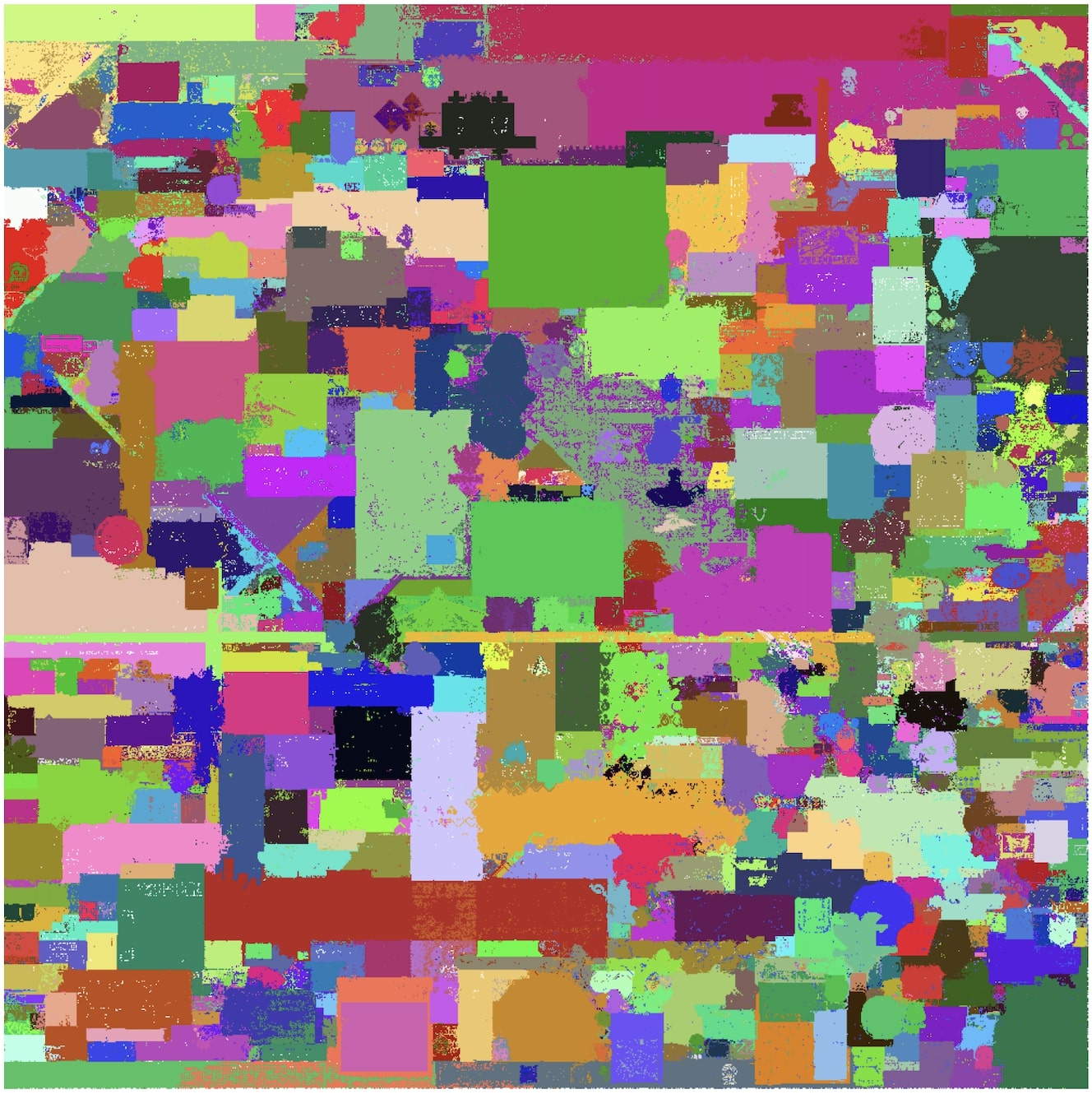}
        \caption{Final clustering}
    \end{subfigure}

    \caption{Canvas snapshots and coalitions discovered at hours 55 and the end of the 2017 experiment.
}
   \label{fig::mid_snapshot_seg}
\end{figure*}

\subsubsection{Coalition Recovery Method}
The approaches described in the previous subsection segment a single snapshot of the r/place canvas. Here, we focus on identifying coalitions across snapshots using dynamic clustering.
Our dynamic clustering algorithm first computes snapshot clusters using the algorithm proposed in the previous section, then combines clusters over time using set cover and a merging scheme.

\textbf{Dynamic clustering via set cover:}  
Given sets $\mathcal{U}$ and $\mathcal{S}$, where all elements in $\mathcal{S}$ are subsets of elements of $\mathcal{U}$, set cover asks for the smallest subset $\mathcal{C} \subseteq \mathcal{S}$ such that $\mathcal{C}$ covers all the elements in $\mathcal{U}$. 
Let $\mathcal{U}$ be the set of all actions during the event, and $\mathcal{S}$ be the set of clusters within snapshots. Each cluster $\mathcal{C}_x \in \mathcal{S}$ contains actions $c_x \in \mathcal{C}_x$ that colors each pixel in the cluster. First, we apply set cover to obtain $\mathcal{C} \subseteq \mathcal{S}$. We then build the set of dynamic clusters by assigning each action to its largest cluster in $\mathcal{C}$. 

\textbf{FGreedy set cover approximation:} The set cover instance resulting from our formulation is expected to have large sets $\mathcal{U}$ and $\mathcal{S}$. However, set cover is NP-hard \cite{Karp1972}. A greedy algorithm that iteratively selects the set covering the most items not covered yet is known to achieve an $O(\log(n))$ approximation \cite{johnson1973approximation} and can be implemented efficiently using the \textit{FGreedy} algorithm \cite{Stergiou-fGreedy}. FGreedy uses a heap to quickly identify the set with the most uncovered items. Nevertheless, the memory requirements of fGreedy due to the large heap size are still prohibitive for our setting. We address this challenge by running multiple iterations of FGreedy constrained to sets with decreasing uncovered size ranges (from large to small sets). At each iteration, our algorithm guarantees that only sets belonging to the greedy solution are selected, which guarantees the same $O(\log(n))$ approximation as the greedy solution.

\textbf{Set cover merging:} The set cover based solution described so far assumes that an artifact has a single action per pixel---as it set belongs to a single snapshot. However, we have noticed that artifacts often have multiple layers due to redundant actions and minor artifact improvements over time. As a result, actions belonging to the same artifact are separated into multiple dynamic clusters. We address this problem by merging set covers based on three notions of their similarity. We associate each set in the set cover solution with its corresponding snapshot cluster. In the first phase, we identify candidate set covers to be merged based on spatial overlap, as those for which the Intersection over Union (IoU) of their snapshot partitions is above a threshold $\alpha_{IoU}$. 
\begin{align*}
    IoU = \frac{\textit{Area of the Intersection of Partitions}}{\textit{Area of the Union of Partitions}}
\end{align*} 
Next, to increase the probability that the partitions correspond to the same artifact, we only merge candidate pairs when the similarity between their areas \textit{AS} satisfies a threshold $\alpha_{AS}$:
\begin{align*}
    AS = \frac{min(\text{Area of Partition One}, \text{Area of Partition Two})}{max(\text{Area of Partition One}, \text{Area of Partition Two})}
\end{align*}
In the second merging phase, the algorithm further merges clusters identified in the previous phase whenever the average embedding of their corresponding participants satisfies a threshold $\alpha_{participant}$. These participant embeddings are the same ones ($M^{participant}$) defined in the previous section. We optimize the thresholds $\alpha_{IoU}$, $\alpha_{AS}$, and $\alpha_{participant}$ manually based on the visual quality of the resulting clusters.




\subsubsection{Experimental Results}


 The experiments were run on a Linux server with an Intel Xeon Gold 6246R 3.4GHz processor (16 cores and 32 threads) with 384GB RAM. We apply the GBIS-N2V-Ward Algorithm to each of the 259,769 snapshots (one per second of the experiment) and obtain $|\mathcal{U}|=$ 1,157,196,961 subsets of the 16,559,897 actions. Clustering all snapshots (in parallel) took approximately 5 days, and it took approximately 10 days to select 1,898,917 sets using the FGreedy algorithm. Finally,  the covers were merged into 39,879 dynamic clusters in approximately 17 hours. 

\begin{figure*}
    \centering
        \subfloat[\textcolor{blue}{\texttt{r/PrequelMemes$^+$}}]{{\includegraphics[width=.35\textwidth]{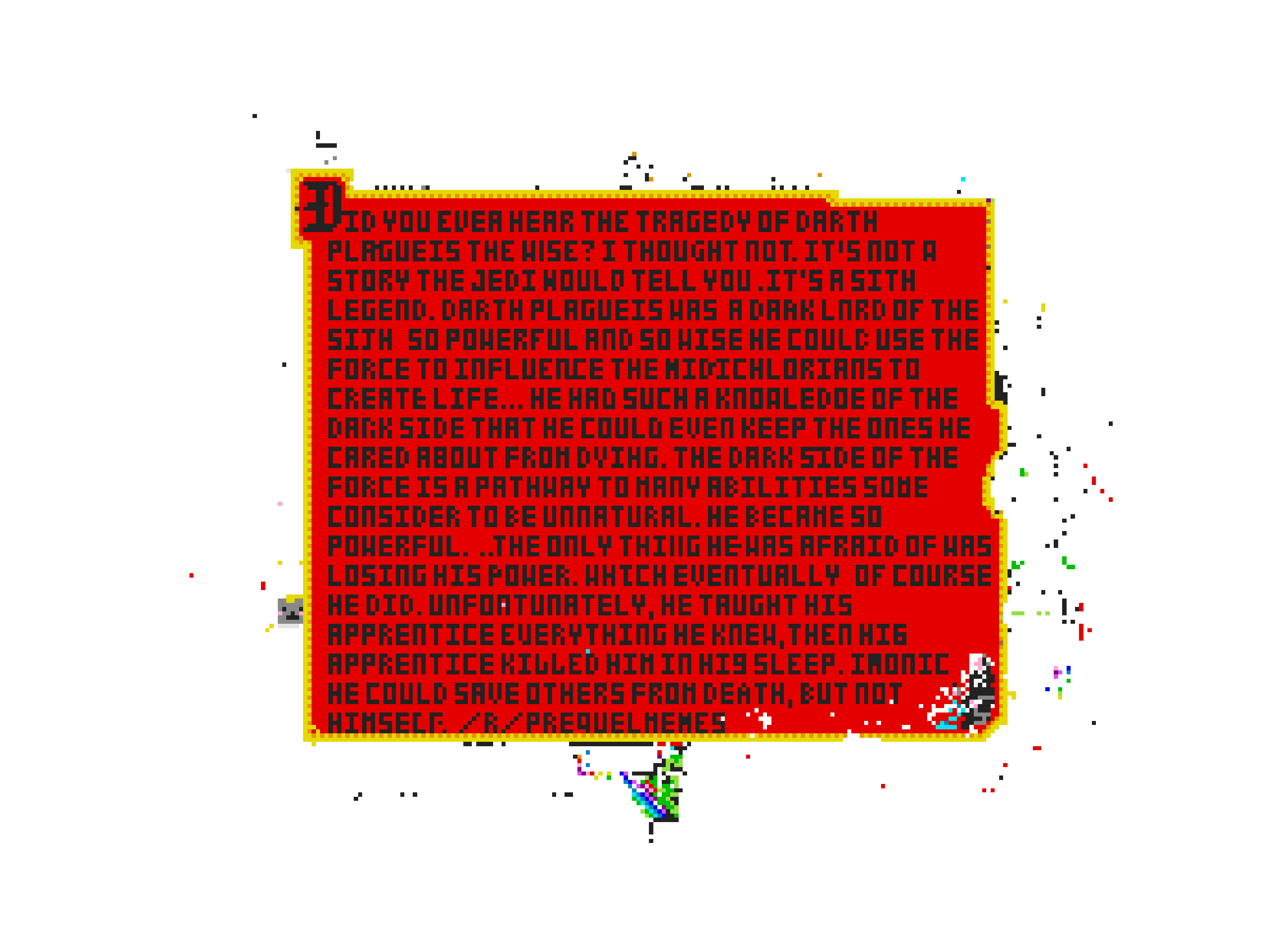}}}
        \subfloat[\texttt{\textcolor{red}{r/PrequelMemes$^-$}}]{{\includegraphics[width=.35\textwidth]{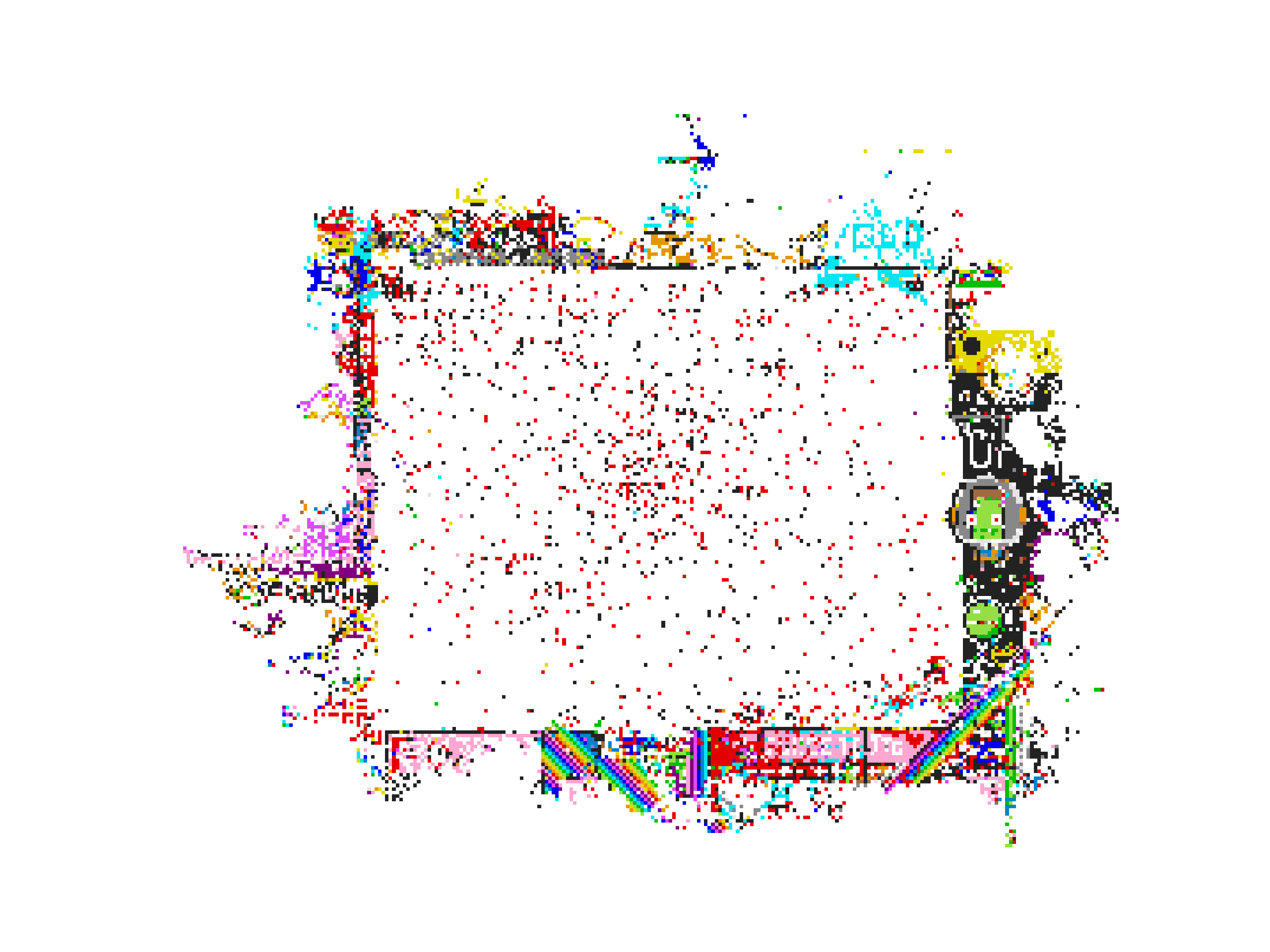}}}

        \subfloat[\textcolor{blue}{\texttt{r/sweden$^+$}}]{{\includegraphics[width=.35\textwidth]{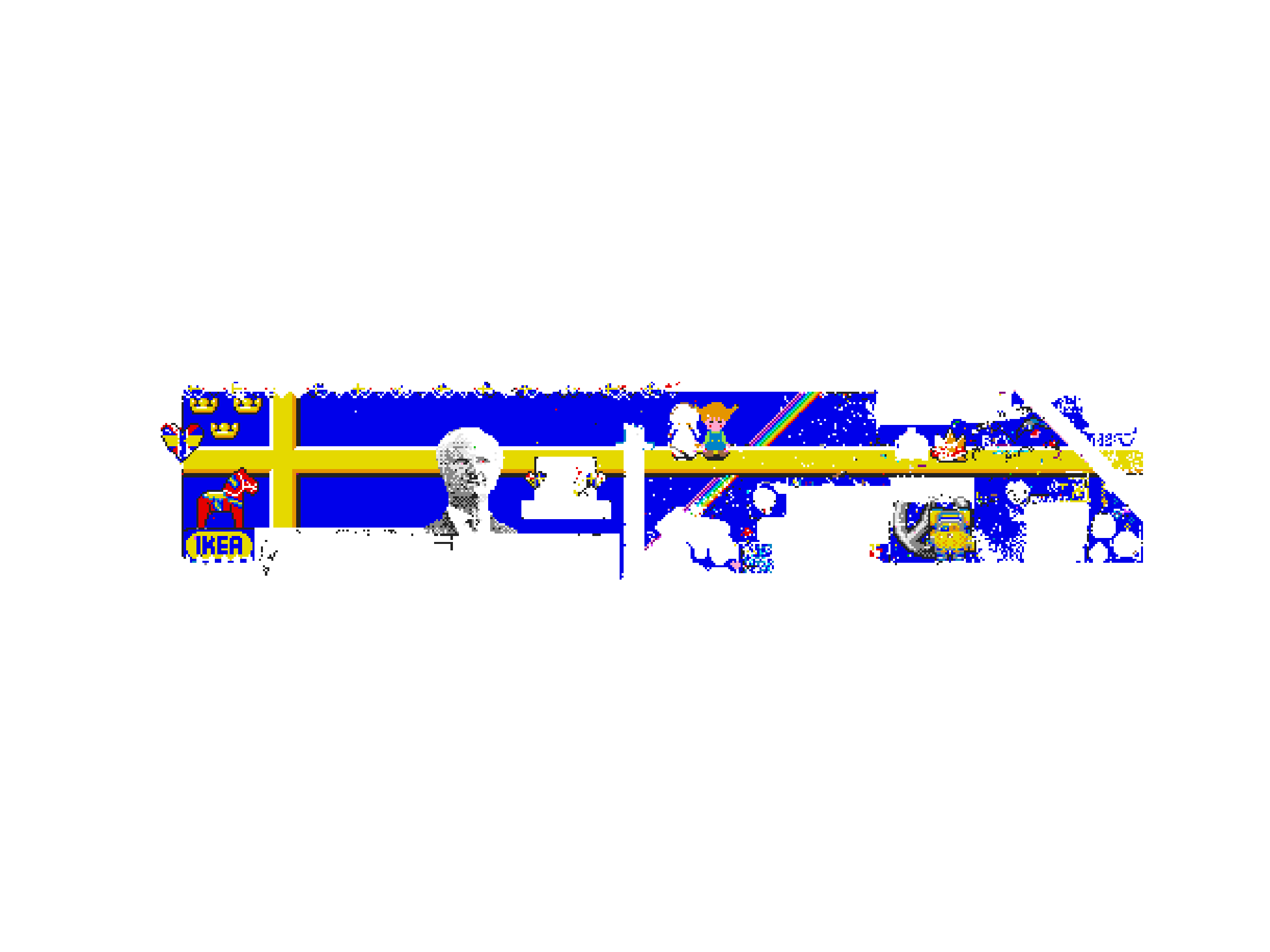}}}
        \subfloat[\texttt{\textcolor{red}{r/sweden$^-$}}]{{\includegraphics[width=.35\textwidth]{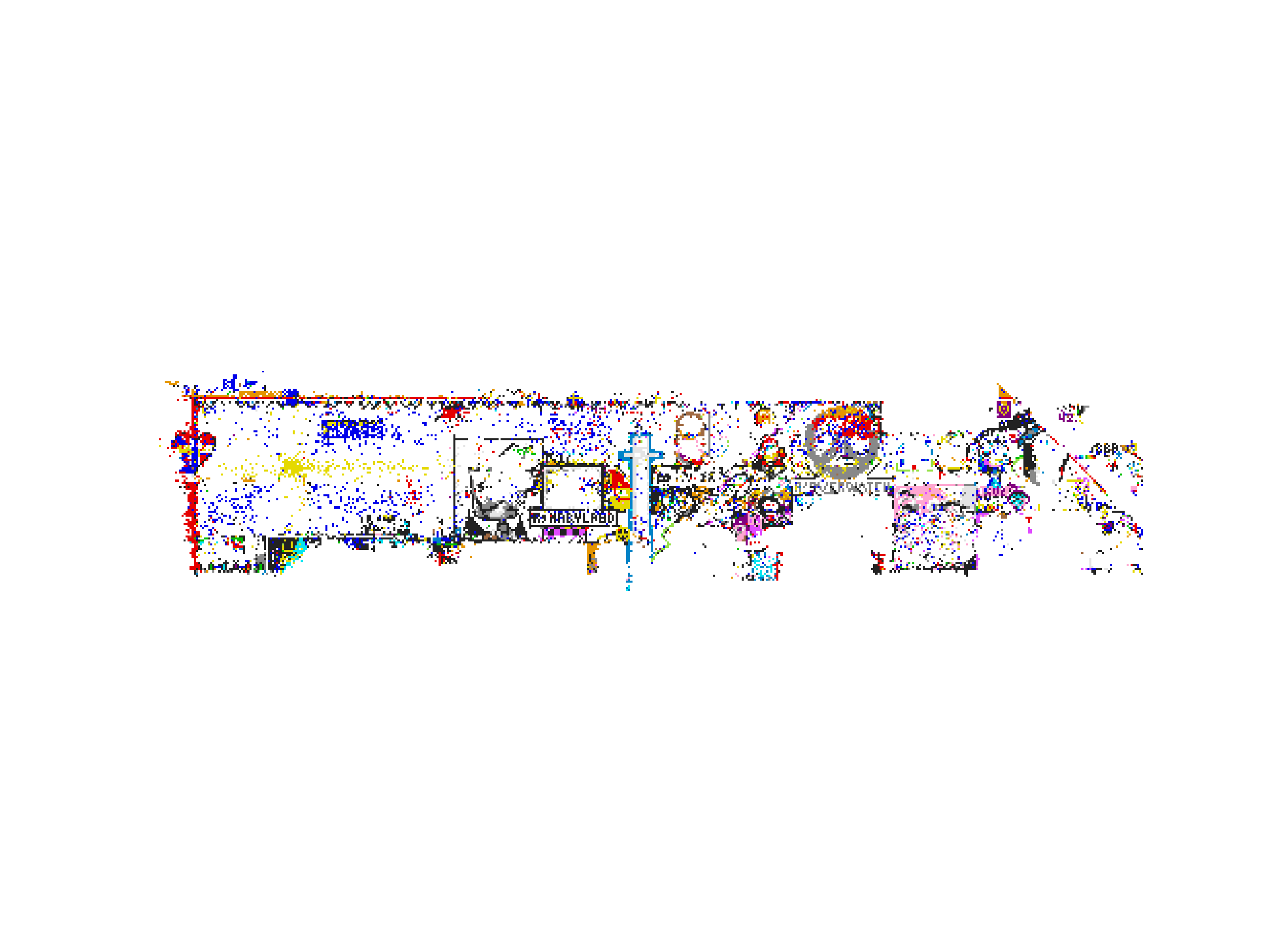}}}

        \subfloat[\texttt{\textcolor{blue}{r/MonaLisaClan$^+$}}]{{\includegraphics[width=.35\textwidth]{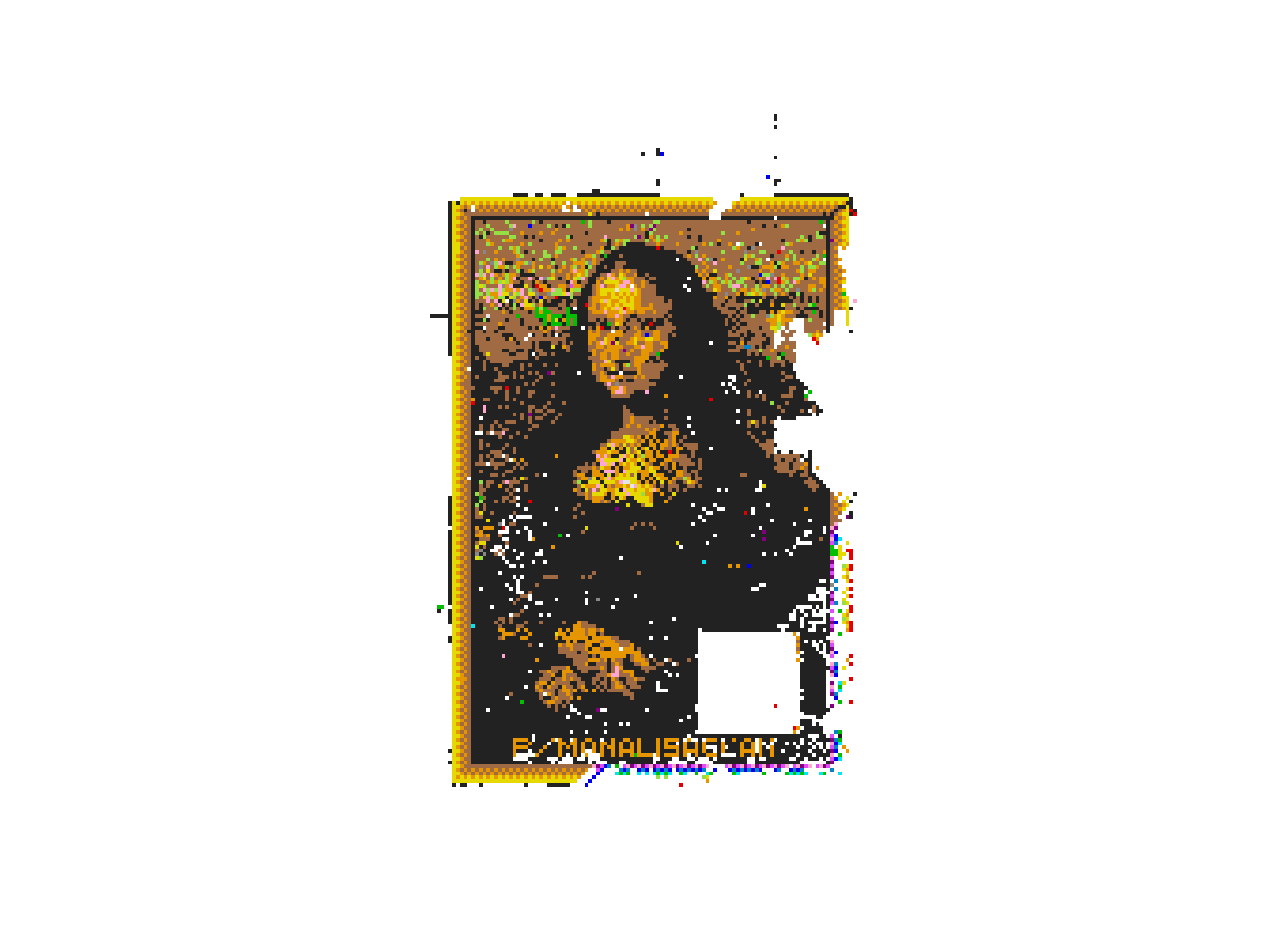}}}
        \subfloat[\texttt{\textcolor{red}{r/MonaLisaClan$^-$}}]{{\includegraphics[width=.35\textwidth]{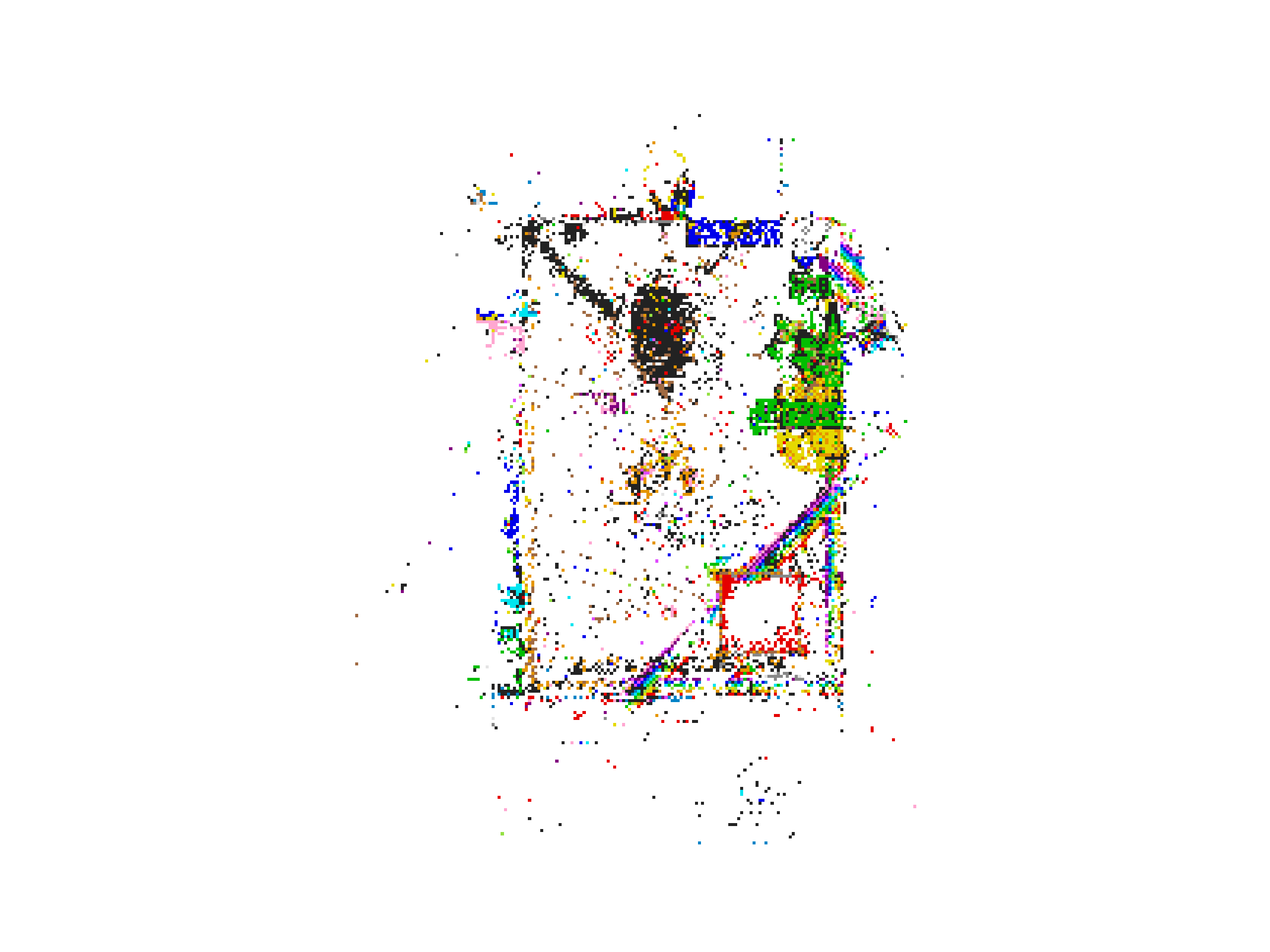}}}

        \subfloat[\texttt{\textcolor{blue}{r/AmericanFlaginPlace$^+$}}]{{\includegraphics[width=.35\textwidth]{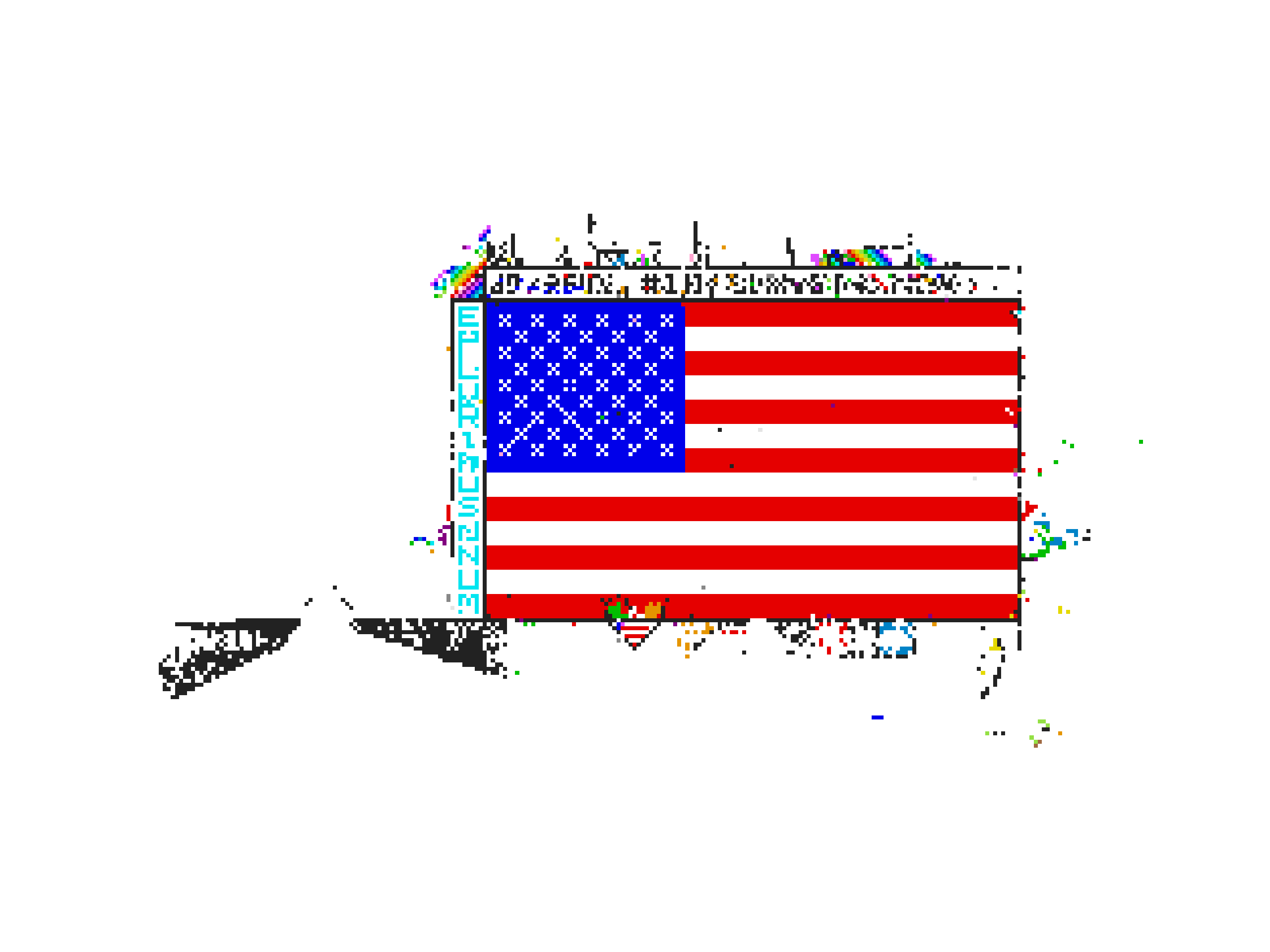}}}
        \subfloat[\textcolor{red}{\texttt{r/AmericanFlaginPlace$^-$}}]{{\includegraphics[width=.35\textwidth]{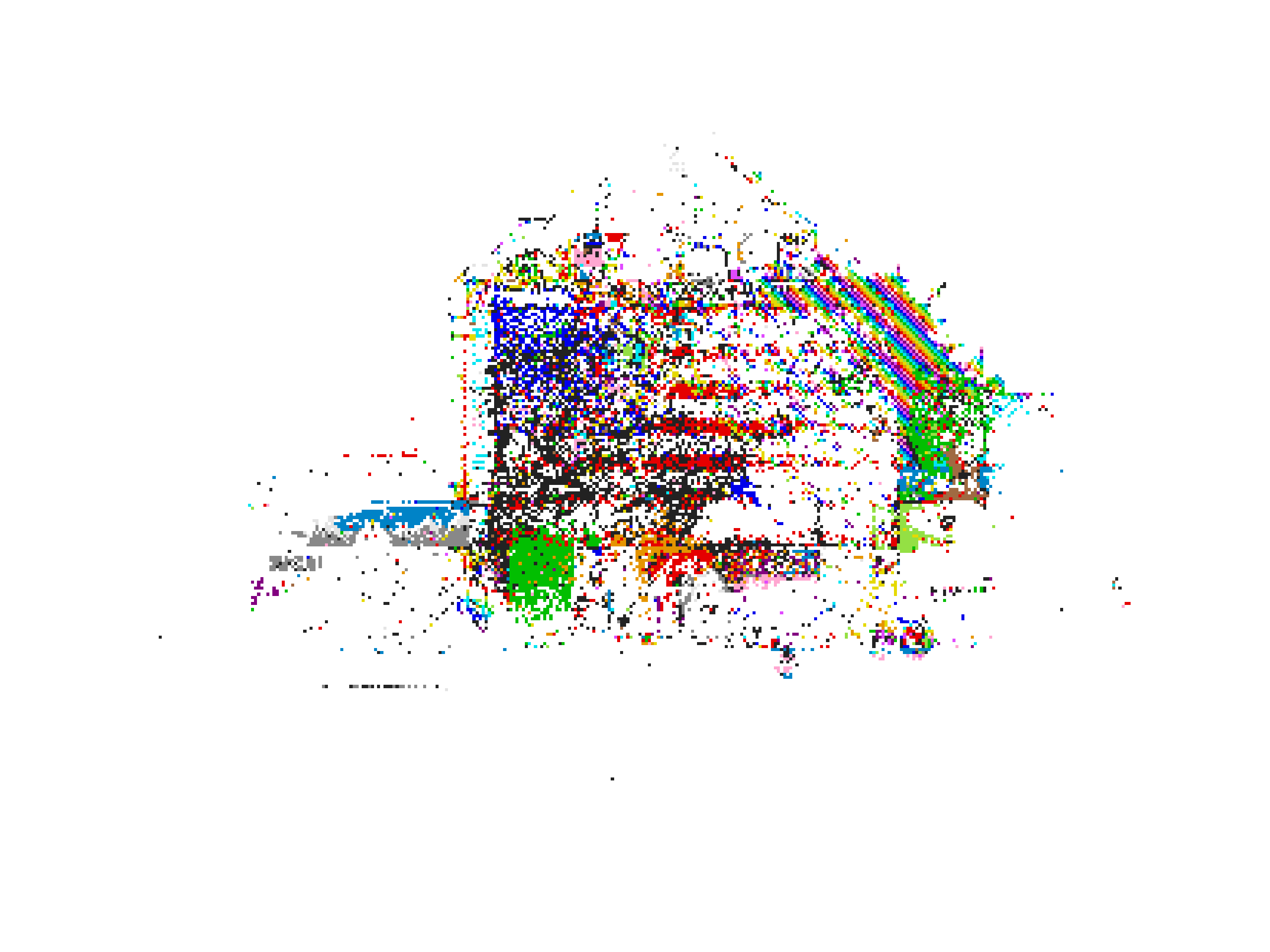}}}
\caption{Examples of successful coalitions with their \textcolor{blue}{collaborative (+)} actions on the left and \textcolor{red}{adversarial (-)} actions on the right. The examples provide additional evidence of the accuracy of the proposed approach, as the large majority of actions are correctly assigned to their corresponding coalition.}
\label{fig::appendix_negative_updates}
\end{figure*}

\textbf{Accuracy}: There are no ground-truth artifacts for the entire duration of the experiment. Thus, we evaluate our solution based on the overlap between the clusters discovered and the smaller set of labeled artifacts. Figure \ref{fig::mid_snapshot_seg} shows the canvas snapshots and the corresponding artifacts discovered at three stages of the experiment. The visualization enables the inspection of the proposed coalition discovery algorithm, further supporting its accuracy beyond the final canvas snapshot. For the final canvas, we apply Adjusted Rand Score (ARS) and Variation of Information (VI)---as in the previous section---to compare the discovered and ground-truth artifacts. The proposed approach achieves an ARS of 0.42 and a VI of 1.96. The results are slightly worse than those presented in Table \ref{tab:experiment_results} because clustering actions over time is more challenging. To the best of our knowledge, there are no existing dynamic clustering algorithms that can cluster images on a canvas over time. These results demonstrate the effectiveness of the proposed clustering algorithm in identifying artifacts throughout the r/place experiment by grouping 16M actions based on visual and participant features. Figure \ref{fig::appendix_negative_updates} shows examples of individual artifacts associated with discovered coalitions, together with their adversarial updates. The results show that the large majority of the actions are correctly labeled.

\subsection{Decision Tree Classification with Various Thresholds (Uncertainty)}

\begin{figure}[h!]
    \centering
        \includegraphics[width=0.5\textwidth]{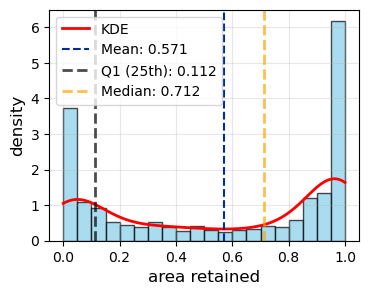}

    \caption{KDE plot of the proportion between the area of artifacts at the final snapshot and the maximum area observed during the game. The dotted lines show the 25th percentile, median, and mean of the distribution.
    }
   \label{fig::area_retain}
\end{figure}

Figure \ref{fig::area_retain} shows the Kernel Density Estimation (KDE) of the ratio between the cluster area at the final snapshot (ground-truth) and the maximum area observed during the experiment. 
In Table \ref{tab:class_result}, we have shown uncertainty results using a success threshold of 11\% of the maximum area of the drawing (25th percentile). 
Tables \ref{tab:class_result_71} and \ref{tab:class_result_57} show uncertainty results for two other success thresholds. The uncertainty is even higher for larger thresholds.

\begin{table}
    \centering
\begin{tabular}{c|c|c|c}\toprule
{Time Period} &  {F1} & {PR AUC} & {Successful Entries ($\%$)}\\  
\midrule
All& 0.14& 0.47&11.48\\
0 - 48 hrs & 0.05 & 0.40 &  7.34\\
48 hrs - end & 0.33 & 0.50 &  23.41 \\
\bottomrule
\end{tabular}
\caption{Coalition success uncertainty in 2017 based on the Decision Tree classifier with 71\% success threshold.
 }\label{tab:class_result_71}  
\end{table}

\begin{table}
    \centering
\begin{tabular}{c|c|c|c}\toprule
{Time Period} &  {F1} & {PR AUC} & {Successful Entries ($\%$)}\\  
\midrule
All& 0.14& 0.46&11.87\\
0 - 48 hrs & 0.06 & 0.41 &  8.03\\
48 hrs - end & 0.32 & 0.49 &  24.09 \\
\bottomrule
\end{tabular}
\caption{Coalition success uncertainty in 2017 based on the Decision Tree classifier with 57\% success threshold.  
 }\label{tab:class_result_57}  
\end{table}

%% file: sample-base.bib
@inproceedings{johnson1973approximation,
  title={Approximation algorithms for combinatorial problems},
  author={Johnson, David S},
  booktitle={STOC},
  year={1973}
}

@misc{simpson-how,
    author = {Simpson, Brian and Lee, Matt and Ellis, Daniel},
    title={How We Built r/Place},
    year = {2017},
    howpublished = {\url{https://www.redditinc.com/blog/how-we-built-rplace/}},
}

@inproceedings{borra2015societal,
  title={Societal controversies in Wikipedia articles},
  author={Borra, Erik and Weltevrede, Esther and Ciuccarelli, Paolo and Kaltenbrunner, Andreas and Laniado, David and Magni, Giovanni and Mauri, Michele and Rogers, Richard and Venturini, Tommaso},
  booktitle={CHI},
  year={2015}
}

@misc{majd-placestart,
    author = {Jordan, Majd and others},
    title={PlaceStart: The Bot that Helped the PlaceStart Team to Preserve its Area},
    year = {2017},
    howpublished = {\url{https://github.com/PlaceStart/placestart}}
}

@misc{stefano-atlas,
    author = {Haagmans, Stefano},
    title={Place Atlas},
    year = {2023},
    howpublished = {\url{https://place-atlas.stefanocoding.me}}
}

@misc{eudaly-fans,
    author = {Eudaly, Zack},
    title={Fans React as xQc Discovers that r/Place is Full of Bots},
    year = {2022},
    howpublished = {\url{https://www.sportskeeda.com/esports/fans-react-xqc-discovers-r-place-full-bots}}
}

@misc{cuthbertson-place,
    author = {Cuthbertson, Anthony},
    title={Reddit Place: The Internet's Best Experiment Yet},
    year = {2017},
    howpublished = {\url{https://www.newsweek.com/reddit-place-internet-experiment-579049}}
}

@misc{lorenz-place,
    author = {Lorenz, Taylor},
    title={Internet Communities are Battling over Pixels},
    year = {2023},
    howpublished = {\url{https://www.washingtonpost.com/technology/2022/04/04/reddit-place-internet-communities/}}
}

@misc{jay-place,
    author = {Peters, Jay},
    title={Reddit’s r/Place is going about as well as expected},
    year = {2023},
    howpublished = {\url{https://www.theverge.com/2023/7/20/23801716/reddits-r-place-protest-art}}
}

@misc{jay2-place,
    author = {Peters, Jay},
    title={Reddit is bringing back r/Place at perhaps the worst possible time},
    year = {2023},
    howpublished = {\url{https://www.theverge.com/2023/7/19/23800309/reddit-r-place-2023-protest}}
}

@misc{jay3-place,
    author = {Peters, Jay},
    title={Reddit expanded the r/Place canvas, and users immediately wrote messages cursing the CEO},
    year = {2023},
    howpublished = {\url{https://www.theverge.com/2023/7/21/23803112/reddit-r-place-canvas-expand-protest-messages-cursing-ceo}}
}

@misc{jody-removal,
    author = {Serrano, Jody},
    title={Reddit Removes Community Drawing of Its CEO Under a Guillotine},
    year = {2023},
    howpublished = {\url{https://themessenger.com/tech/reddit-removes-community-drawing-of-its-ceo-under-a-guillotine}}
}

@inproceedings{rappaz-place,
author = {Rappaz, Jérémie and Catasta, Michele and West, Robert and Aberer, Karl},
year = {2018},
title = {Latent Structure in Collaboration: The Case of Reddit r/place},
booktitle= {ICWSM},
}

@article{chen2021collaborative,
  title={Collaborative learning at scale},
  author={Chen, Bodong and H{\aa}klev, Stian and Ros{\'e}, Carolyn Penstein},
  journal={International Handbook of Computer-supported Collaborative Learning},
  pages={163--181},
  year={2021},
  publisher={Springer}
}

@inproceedings{vachher2020understanding,
  title={Understanding community-level conflicts through Reddit r/place},
  author={Vachher, Prateek and Levonian, Zachary and Cheng, Hao-Fei and Yarosh, Svetlana},
  booktitle={CSCW},
  year={2020}
}

@inproceedings{israeli2022must,
  title={This Must Be the Place: Predicting Engagement of Online Communities in a Large-scale Distributed Campaign},
  author={Israeli, Abraham and Kremiansky, Alexander and Tsur, Oren},
  booktitle={WebConf},
  year={2022}
}

@article{pendergrass2022digital,
  title={Digital mandalas: Communication and authentic human interaction in reddit's r/place platform.},
  author={Pendergrass, William and Compomizzi, Joseph and Scibelli, David and Szarmach, Matthew},
  journal={Issues in Information Systems},
  volume={23},
  number={3},
  year={2022}
}

@article{cummings_07,
title = {Coordination costs and project outcomes in multi-university collaborations},
  author={Cummings, Jonathan N. and Kiesler, Sara},

journal = {Research Policy},
volume = {36},
number = {10},
pages = {1620-1634},
year = {2007},
issn = {0048-7333}
}

@article{Pendharkar_09,
author = {Pendharkar, Parag C. and Rodger, James A.},
title = {The relationship between software development team size and software development cost},
year = {2009},
publisher = {Association for Computing Machinery},
address = {New York, NY, USA},
volume = {52},
number = {1},
issn = {0001-0782},
journal = {CACM},
pages = {141–144},
numpages = {4}
}

@article{litherland2021instruction,
  title={Instruction vs. emergence on r/place: Understanding the growth and control of evolving artifacts in mass collaboration},
  author={Litherland, Kristina T and M{\o}rch, Anders I},
  journal={Computers in Human Behavior},
  volume={122},
  pages={106845},
  year={2021},
  publisher={Elsevier}
}

@article{kapoor_lee,
 ISSN = {01432095, 10970266},
 author = {Rahul Kapoor and Joon Mahn Lee},
 journal = {Strategic Management Journal},
 number = {3},
 pages = {274--296},
 publisher = {Wiley},
 title = {Coordinating and competing in ecosystems: How organizational forms shape new technology investments},
 urldate = {2025-04-10},
 volume = {34},
 year = {2013}
}

@article{Shimrat1962,
 author = {Shimrat, M.},
 title = {Algorithm 112: Position of Point Relative to Polygon},
 journal = {CACM},
year = {1962},
 volume = {5},
 number = {8}
}

@inproceedings{kwak2010twitter,
  title={What is Twitter, a social network or a news media?},
  author={Kwak, Haewoon and Lee, Changhyun and Park, Hosung and Moon, Sue},
  booktitle={WebConf},
  year={2010}
}

@inproceedings{leskovec2009meme,
  title={Meme-tracking and the dynamics of the news cycle},
  author={Leskovec, Jure and Backstrom, Lars and Kleinberg, Jon},
  booktitle={SIGKDD},
  year={2009}
}

@inproceedings{cheng2019makes,
  title={What makes a good team? a large-scale study on the effect of team composition in honor of kings},
  author={Cheng, Ziqiang and Yang, Yang and Tan, Chenhao and Cheng, Denny and Cheng, Alex and Zhuang, Yueting},
  booktitle={WebConf},
  year={2019}
}

@inproceedings{gupta2022instagram,
  title={Instagram of Rivers: Facilitating Distributed Collaboration in Hyperlocal Citizen Science},
  author={Gupta, Srishti and Jablonski, Julia and Tsai, Chun-Hua and Carroll, John M},
  booktitle={CSCW},
  year={2022},
}

@inproceedings{jagannath2020we,
  title={"(We) Can Talk It Out...": Designing for Promoting Conflict-Resolution Skills in Youth on a Moderated Minecraft Server},
  author={Jagannath, Krithika and Salen, Katie and Slov{\`a}k, Petr},
  booktitle={CSCW},
  year={2020},
}

@article{Felzenswalb-graphcut,
author = {Felzenszwalb, Pedro and Huttenlocher, Daniel},
year = {2004},
month = {09},
pages = {167-181},
title = {Efficient Graph-Based Image Segmentation},
volume = {59},
journal = {IJCV},
}

@article{jain1999data,
  title={Data clustering: a review},
  author={Jain, Anil K and Murty, M Narasimha and Flynn, Patrick J},
  journal={CSUR},
  volume={31},
  number={3},
  pages={264--323},
  year={1999},
  publisher={Acm New York, NY, USA}
}

@article{fortunato2010community,
  title={Community detection in graphs},
  author={Fortunato, Santo},
  journal={Physics reports},
  volume={486},
  number={3-5},
  pages={75--174},
  year={2010},
}

@article{newman2006modularity,
  title={Modularity and community structure in networks},
  author={Newman, Mark EJ},
  journal={PNAS},
  volume={103},
  number={23},
  pages={8577--8582},
  year={2006},
  publisher={National Acad Sciences}
}

@article{ward1963hierarchical,
  title={Hierarchical grouping to optimize an objective function},
  author={Ward Jr, Joe H},
  journal={JASA},
  volume={58},
  number={301},
  pages={236--244},
  year={1963},
  publisher={Taylor \& Francis}
}

@article{muller2018compression,
  title={Compression in cultural evolution: Homogeneity and structure in the emergence and evolution of a large-scale online collaborative art project},
  author={M{\"u}ller, Thomas F and Winters, James},
  journal={PloS one},
  volume={13},
  number={9},
  pages={e0202019},
  year={2018},
  publisher={Public Library of Science San Francisco, CA USA}
}

@mastersthesis{armstrong2018coordination,
  title={Coordination in a Peer Production Platform: A study of Reddit's/r/Place experiment},
  author={Armstrong, Ben},
  year={2018},
  school={University of Waterloo}
}

@article{mason2012collaborative,
  title={Collaborative learning in networks},
  author={Mason, Winter and Watts, Duncan J},
  journal={PNAS},
  volume={109},
  number={3},
  pages={764--769},
  year={2012},
  publisher={National Acad Sciences}
}

@article{malone2010collective,
  title={The collective intelligence genome},
  author={Malone, Thomas W and Laubacher, Robert and Dellarocas, Chrysanthos},
  journal={MIT SMR},
   volume={51},
  number={3},
  pages={21},
  year={2010},
  publisher={MIT}
}

@article{rand2013human,
  title={Human cooperation},
  author={Rand, David G and Nowak, Martin A},
  journal={Trends in cognitive sciences},
  volume={17},
  number={8},
  pages={413--425},
  year={2013},
}

@article{israeli2023flying,
  title={With Flying Colors: Predicting Community Success in Large-scale Collaborative Campaigns},
  author={Israeli, Abraham and Tsur, Oren},
  journal={arXiv preprint arXiv:2307.09650},
  year={2023}
}

@inproceedings{grover2016node2vec,
  title={node2vec: Scalable feature learning for networks},
  author={Grover, Aditya and Leskovec, Jure},
  booktitle={SIGKDD},
  year={2016}
}

@inproceedings{meilua2003comparing,
  title={Comparing clusterings by the variation of information},
  author={Meil{\u{a}}, Marina},
  booktitle={COLT},
  year={2003},
}

@article{hubert1985comparing,
  title={Comparing partitions},
  author={Hubert, Lawrence and Arabie, Phipps},
  journal={Journal of classification},
  volume={2},
  pages={193--218},
  year={1985},
  publisher={Springer}
}

@inproceedings{Stergiou-fGreedy,
author = {Stergiou, Stergios and Tsioutsiouliklis, Kostas},
title = {Set Cover at Web Scale},
year = {2015},
booktitle = {SIGKDD},
}

@Inbook{Karp1972,
author="Karp, Richard M.",
editor="Miller, Raymond E.
and Thatcher, James W.
and Bohlinger, Jean D.",
title="Reducibility among Combinatorial Problems",
bookTitle="Complexity of Computer Computations: Proceedings of a symposium on the Complexity of Computer Computations, held March 20--22, 1972, at the IBM Thomas J. Watson Research Center, Yorktown Heights, New York, and sponsored by the Office of Naval Research, Mathematics Program, IBM World Trade Corporation, and the IBM Research Mathematical Sciences Department",
year="1972",
publisher="Springer US",
address="Boston, MA",
pages="85--103",
isbn="978-1-4684-2001-2",
}

@article{ingham1974ringelmann,
  title={The Ringelmann effect: Studies of group size and group performance},
  author={Ingham, Alan G and Levinger, George and Graves, James and Peckham, Vaughn},
  journal={Journal of Experimental Social Psychology},
  volume={10},
  number={4},
  pages={371--384},
  year={1974},
}

@article{richerson2016cultural,
  title={Cultural group selection plays an essential role in explaining human cooperation: A sketch of the evidence},
  author={Richerson, Peter and Baldini, Ryan and Bell, Adrian V and Demps, Kathryn and Frost, Karl and Hillis, Vicken and Mathew, Sarah and Newton, Emily K and Naar, Nicole and Newson, Lesley and others},
  journal={Behavioral and Brain Sciences},
  volume={39},
  pages={e30},
  year={2016},
  publisher={Cambridge University Press}
}

@article{tajfel2001integrative,
  title={An integrative theory of intergroup conflict},
  author={Tajfel, Henri and Turner, John and Austin, William G and Worchel, Stephen},
  journal={Intergroup relations: Essential readings},
  pages={94--109},
  year={2001}
}

@inproceedings{campbell1965ethnocentric,
  title={Ethnocentric and other altruistic motives},
  author={Campbell, Donald T},
  booktitle={Nebraska symposium on motivation},
  volume={13},
  pages={283},
  year={1965}
}

@article{chen2008player,
  title={Player guild dynamics and evolution in massively multiplayer online games},
  author={Chen, Chien-Hsun and Sun, Chuen-Tsai and Hsieh, Jilung},
  journal={CyberPsychology \& Behavior},
  volume={11},
  number={3},
  pages={293--301},
  year={2008},
  publisher={Mary Ann Liebert}
}

@article{benkler2002coase,
  title={Coase's penguin, or, Linux and" The nature of the firm"},
  author={Benkler, Yochai},
  journal={Yale law journal},
  pages={369--446},
  year={2002},
  publisher={JSTOR}
}

@inproceedings{forte2008scaling,
  title={Scaling consensus: Increasing decentralization in Wikipedia governance},
  author={Forte, Andrea and Bruckman, Amy},
  booktitle={Proceedings of the 41st Annual Hawaii International Conference on System Sciences (HICSS 2008)},
  pages={157--157},
  year={2008},
  organization={IEEE}
}

@book{benkler2006wealth,
  title={The Wealth of Networks: How Social Production Transforms Markets and Freedom},
  author={Benkler, Yochai},
  year={2006},
  publisher={Yale University Press}
}

@article{rafaeli2008online,
  title={Online motivational factors: Incentives for participation and contribution in Wikipedia},
  author={Rafaeli, Sheizaf and Ariel, Yaron},
  journal={Psychological aspects of cyberspace: Theory, research, applications},
  volume={2},
  number={08},
  pages={243--267},
  year={2008}
}

@article{forte2009decentralization,
  title={Decentralization in Wikipedia governance},
  author={Forte, Andrea and Larco, Vanesa and Bruckman, Amy},
  journal={Journal of Management Information Systems},
  volume={26},
  number={1},
  pages={49--72},
  year={2009},
  publisher={Taylor \& Francis}
}

@book{kraut2012building,
  title={Building successful online communities: Evidence-based social design},
  author={Kraut, Robert E and Resnick, Paul},
  year={2012},
  publisher={MIT Press},
  address={Cambridge, MA, USA}
}

@inproceedings{kittur2008harnessing,
  title={Harnessing the wisdom of crowds in {Wikipedia}: quality through coordination},
  author={Kittur, Aniket and Kraut, Robert E},
  booktitle={Proceedings of the 2008 ACM conference on Computer supported cooperative work},
  pages={37--46},
  year={2008},
  publisher={ACM},
  doi={10.1145/1460563.1460572}
}

@article{lang2022outgroup,
  title={Outgroup threat and the emergence of cohesive groups: A cross-cultural examination},
  author={Lang, Martin and Xygalatas, Dimitris and Kavanagh, Christopher M and Craciun Boccardi, Natalia A and Halberstadt, Jamin and Jackson, Chris and Mart{\'\i}nez, Mercedes and Reddish, Paul and Tong, Eddie MW and V{\'a}zquez, Alexandra and others},
  journal={Group Processes \& Intergroup Relations},
  volume={25},
  number={7},
  pages={1739--1759},
  year={2022},
  publisher={Sage Publications Sage UK: London, England}
}

@inproceedings{marlow2013impression,
  title={Impression formation in online peer production: activity traces and personal profiles in github},
  author={Marlow, Jennifer and Dabbish, Laura and Herbsleb, Jim},
  booktitle={Proceedings of the 2013 conference on Computer supported cooperative work},
  pages={117--128},
  year={2013}
}

@article{postmes2002collective,
  title={Collective action in the age of the Internet: Mass communication and online mobilization},
  author={Postmes, Tom and Brunsting, Suzanne},
  journal={Social science computer review},
  volume={20},
  number={3},
  pages={290--301},
  year={2002},
  publisher={Sage Publications Sage CA: Thousand Oaks, CA}
}

@inproceedings{raban2010empirical,
  title={An empirical study of critical mass and online community survival},
  author={Raban, Daphne R and Moldovan, Mihai and Jones, Quentin},
  booktitle={Proceedings of the 2010 ACM conference on Computer supported cooperative work},
  pages={71--80},
  year={2010}
}

@article{turner2005picturing,
  title={Picturing Usenet: Mapping computer-mediated collective action},
  author={Turner, Tammara Combs and Smith, Marc A and Fisher, Danyel and Welser, Howard T},
  journal={Journal of Computer-Mediated Communication},
  volume={10},
  number={4},
  pages={JCMC1048},
  year={2005},
  publisher={Oxford University Press Oxford, UK}
}

@article{iriberri2009life,
  title={A life-cycle perspective on online community success},
  author={Iriberri, Alicia and Leroy, Gondy},
  journal={ACM Computing Surveys (CSUR)},
  volume={41},
  number={2},
  pages={1--29},
  year={2009},
  publisher={Acm New York, NY, USA}
}

@article{anand2023game,
  title={A game-theoretic analysis of Wikipedia’s peer production: The interplay between community’s governance and contributors’ interactions},
  author={Anand, Santhanakrishnan and Arazy, Ofer and Mandayam, Narayan and Nov, Oded},
  journal={PLOS ONE},
  volume={18},
  number={5},
  pages={1--31},
  year={2023},
  publisher={Public Library of Science}
}

@inproceedings{kriplean2007community,
  title={Community, consensus, coercion, control: cs* w or how policy mediates mass participation},
  author={Kriplean, Travis and Beschastnikh, Ivan and McDonald, David W and Golder, Scott A},
  booktitle={Proceedings of the 2007 ACM International Conference on Supporting Group Work},
  pages={167--176},
  year={2007}
}

@inproceedings{zhang2017community,
  title={Community identity and user engagement in a multi-community landscape},
  author={Zhang, Justine and Hamilton, William and Danescu-Niculescu-Mizil, Cristian and Jurafsky, Dan and Leskovec, Jure},
  booktitle={Proceedings of the international AAAI conference on web and social media},
  volume={11},
  number={1},
  pages={377--386},
  year={2017}
}

@article{halfaker2013rise,
  title={The rise and decline of an open collaboration system: How Wikipedia’s reaction to popularity is causing its decline},
  author={Halfaker, Aaron and Geiger, R Stuart and Morgan, Jonathan T and Riedl, John},
  journal={American behavioral scientist},
  volume={57},
  number={5},
  pages={664--688},
  year={2013},
  publisher={SAGE Publications Sage CA: Los Angeles, CA}
}

@inproceedings{ortega2008inequality,
  title={On the inequality of contributions to Wikipedia},
  author={Ortega, Felipe and Gonzalez-Barahona, Jesus M and Robles, Gregorio},
  booktitle={Proceedings of the 41st Annual Hawaii International Conference on System Sciences (HICSS 2008)},
  pages={304--304},
  year={2008},
  organization={IEEE}
}
